\documentclass[fleqn,usenatbib]{mnras}

\usepackage{newtxtext,newtxmath}
\usepackage[dvipsnames]{xcolor}
\usepackage{soul}

\usepackage[T1]{fontenc}
\usepackage{ae,aecompl}

\usepackage{graphicx}	
\usepackage{amsmath}	

\newcommand{\galf}[0]{{\tt GALFORM} }

\usepackage{subfig}

\title[]{The PAU Survey: a new constraint on galaxy formation models using the observed colour redshift relation. }

\author[G. Manzoni et al.]{%
G.~Manzoni$^{1,2,3,4}$\thanks{E-mail: iasmanzoni@ust.hk}, %
C.~M.~Baugh$^{2,4}$, %
P.~Norberg$^{2,3,4}$, %
L.~Cabayol$^{5,6}$, %
J.~L.~van den Busch$^{7}$, %
A.~Wittje$^{7}$, %
\newauthor%
D.~Navarro-Giron\'es$^{8,9}$, %
M.~Eriksen$^{5,6}$,
P.~Fosalba$^{8,9}$, 
J.~Carretero$^{5,6}$,
F.~J.~Castander$^{8,9}$
R.~Casas$^{8,9}$,
\newauthor
J.~De~Vicente$^{11}$,
E.~Fernandez$^{5}$,
J.~Garc\'ia-Bellido$^{12}$,
E.~Gaztanaga$^{10,8}$,
J.~C.~Helly$^{2}$,
H.~Hoekstra$^{13}$,
\newauthor
H.~Hildebrandt$^{7}$,
E.~J.~Gonzalez$^{5,14,6}$,
S.~Koonkor$^{2,4}$,
R.~Miquel$^{5,15}$,
C.~Padilla$^{5}$,
P.~Renard$^{16}$,
\newauthor
E.~Sanchez$^{11}$,
I.~Sevilla-Noarbe$^{11}$,
M.~Siudek$^{8}$,
J.~Y.~H.~Soo$^{17}$,
P.~Tallada-Cresp\`i$^{11,6}$,
L.~Tortorelli$^{18}$\\ \\
$^{1}$Jockey Club Institute for Advanced Study, The Hong Kong University of Science and Technology, Hong Kong S.A.R., China\\
$^{2}$Institute for Computational Cosmology (ICC), Department of Physics, Durham University, South Road, Durham DH1 3LE, UK\\
$^{3}$Centre for Extragalactic Astronomy (CEA), Department of Physics, Durham University, South Road, Durham DH1 3LE, UK\\
$^{4}$Institute for Data Science, Durham University, South Road, Durham DH1 3LE, UK\\
$^{5}$Institut de F\'isica d’Altes Energies (IFAE), The Barcelona Institute of Science and Technology, Campus UAB, 08193 Bellaterra (Barcelona), Spain.\\
$^{6}$Port d’Informaci\'o Cient\'ifica (PIC), Campus UAB, C. Albareda s/n, 08193 Bellaterra (Barcelona), Spain.\\
$^{7}$Ruhr University Bochum, Faculty of Physics and Astronomy, Astronomical Institute (AIRUB), German Centre for Cosmological Lensing, 44780 Bochum, \\
Germany. \\
$^{8}$Institute of Space Sciences (ICE, CSIC), Campus UAB, Carrer de Can Magrans, s/n, 08193 Barcelona, Spain\\
$^{9}$Institut d'Estudis Espacials de Catalunya (IEEC), Gran Capit\`a, 2-4, Edifici Nexus, Desp. 201, 08034 Barcelona, Spain\\
$^{10}$Institute of Cosmology \& Gravitation, University of Portsmouth, Dennis Sciama Building, Burnaby Road, Portsmouth PO1 3FX, UK\\
$^{11}$Centro de Investigaciones Energéticas, Medioambientales y Tecnológicas (CIEMAT), Avenida Complutense 40, E-28040 Madrid (Spain)\\
$^{12}$Instituto de F\'isica Te\'orica UAM-CSIC, Universidad Auton\'oma de Madrid, Cantoblanco 28049 Madrid, Spain\\
$^{13}$Leiden Observatory, Leiden University, PO Box 9513, 2300 RA, Leiden, The Netherlands\\
$^{14}$Instituto de Astronomía Teórica y Experimental (IATE-CONICET), Laprida 854, X5000BGR, C\'ordoba, Argentina.\\
$^{15}$Instituci\`o Catalana de Recerca i Estudis Avan\c{c}ats (ICREA), 08010 Barcelona, Spain\\
$^{16}$Department of Astronomy, Tsinghua University, Beijing 100084, China\\
$^{17}$School of Physics, Universiti Sains Malaysia, 11800 USM, Pulau Pinang, Malaysia\\
$^{18}$University Observatory, Faculty of Physics, Ludwig-Maximilians-Universit\"at M\"unchen, Scheinerstr. 1, 81679 Munich, Germany\\
}

\date{Accepted 2024 March 02. Received 2024 February 29; in original form 2023 November 17}

\pubyear{2024}

\begin{document}
\label{firstpage}
\pagerange{\pageref{firstpage}--\pageref{lastpage}}
\maketitle

\begin{abstract}
We use the \texttt{GALFORM} semi-analytical galaxy formation model implemented in the Planck Millennium N-body simulation to build a mock galaxy catalogue on an observer's past lightcone. The mass resolution of this N-body simulation is almost an order of magnitude better than in previous simulations used for this purpose, allowing us to probe fainter galaxies and hence build a more complete mock catalogue at low redshifts. The high time cadence of the simulation outputs allows us to make improved calculations of galaxy properties and positions in the mock. We test the predictions of the mock against the Physics of the Accelerating Universe Survey, a narrow band imaging survey with highly accurate and precise photometric redshifts, which probes the galaxy population over a lookback time of 8 billion years. We compare the model against the observed number counts, redshift distribution and evolution of the observed colours and find good agreement; these statistics avoid the need for model-dependent processing of the observations. The model produces red and blue populations that have similar median colours to the observations. However, the bimodality of galaxy colours in the model is stronger than in the observations. This bimodality is reduced on including a simple model for errors in the \galf photometry. We examine how the model predictions for the observed galaxy colours change when perturbing key model parameters. This exercise shows that the median colours and relative abundance of red and blue galaxies provide constraints on the strength of the feedback driven by supernovae used in the model. 
\end{abstract}





\section{Introduction}

In the effort to understand the physical processes that govern the formation and evolution of galaxies, mock galaxy catalogues have become an important tool for comparing theoretical models to observations. Wide-field galaxy redshift surveys are covering ever larger areas of the sky to increasing depths. A mock catalogue can be used to model the selection effects that dominate every galaxy survey, and hence allows us to understand how these observational effects shape any measurements made from the survey, and thus, in turn, helps us to disentangle physical results from observational features. 

Here, with the aim of using new observations to help constrain galaxy formation models, we build a replica of the Physics of the Accelerating Universe Survey (PAUS; \citealt{eriksen19, padilla19, Serrano:2023}; Castander et~al.~in prep). 
Using a combination of the PAUS narrow band imaging and intermediate and broad band photometry, \cite{eriksen19} measured photometric redshifts for PAUS galaxies in the COSMOS field, estimating a scatter ($\sigma_{68}/(1 + z) = 0.0037$ to $i_{\rm AB} = 22.5$) that is around an order of magnitude below the few percent level that is typically obtained when using a handful of broad band filters (see also \citealt{Eriksen2020}, \citealt{alarcon21}, \citealt{cabayol21}, \citealt{Soo:2021}, \citealt{Cabayol:2023}, \citealt{david23}).   

Building a mock catalogue with realistic photometric redshift errors provides a way to understand the selection effects on measured statistics. We focus on two of the largest fields in PAUS, the Canada-France-Hawaii-Telescope Lensing Survey (CFHTLS) W1 and W3 fields, which cover about $38$ deg$^2$. Broad band imaging is available for these fields in the standard $u^{*}$, $g$, $r$, $i$, $z$ filter set from the CFHTLenS catalogues \citep{CFHTLS:2012,Erben:2013}, to complement the PAUS narrow band photometry.  
Despite the much improved precision in the photometric redshifts obtained using PAUS photometry, the associated errors along the line-of-sight remain an observational effect of concern. The \textit{rms} error at $z \sim 0.3$ is a little over a comoving distance of $10 h^{-1}$ Mpc \citep{stothert18}. Also, around 17 per cent of the galaxies in the $i_{\rm AB}=22.5$ sample with photometric redshifts have substantial errors in their estimated redshifts and are considered as outliers (see Eq.~\ref{eq:outliers} for the definition of a photometric redshift outlier, which is the one used by \citealt{eriksen19}). Such errors could alter the perceived evolution of a statistic by mixing galaxies with different properties between redshift bins. If the property evolves over a redshift range comparable to the errors in the photometric redshift, or if there are significant numbers of redshift outliers, this will alter the measured evolution of the statistic. The mock catalogue allows us to investigate the impact of errors in photometry and, in turn, photometric redshifts, on observed galaxy statistics. 

The PAU Survey complements and extends spectroscopic studies of galaxy evolution. PAUS is deeper than the Galaxy and Mass Assembly (GAMA) Survey \citep{driver09}. The deepest GAMA fields are limited to $r_{\rm AB}=19.8$. For the typical galaxy colour of $r-i \sim 0.4$ \citep{gonzalez09}, this corresponds roughly to $i_{\rm AB}=20.2$, which is approximately two magnitudes shallower than the PAUS limit considered here of $i_{\rm AB}=22.5$. (Note that the PAUS catalogue now extends to $i_{\rm AB}=23$, but when this project was started the bulk of the available photometry was limited to $i_{\rm AB}=22.5$.)
The GAMA redshift distribution peaks at $z \sim 0.2$ with a tail that extends to $z \sim 0.5$. PAUS has the same depth as the VIMOS Public Extragalactic Redshift Survey (VIPERS; \citealt{guzzo14,scodeggio18}), which measured approximately 100\,000 galaxy redshifts in the interval $0.5<z<1.2$, over 24 deg$^2$, around two-thirds of the combined solid angle of the W1 and W3 fields considered here. VIPERS used a colour preselection to target galaxies with $z\gtrsim0.5$. As we will see, this is the peak in the redshift distribution for galaxies brighter than $i_{\rm AB}=22.5$. 
The Deep Extragalactic VIsible Legacy Survey (DEVILS) \cite{Davies:2018} is deeper than GAMA with a higher completeness than surveys like VIPERS, but covers a small solid angle (6 square degrees) and contains $60 \, 000$ galaxies. PAUS samples the full range of galaxy redshifts to this magnitude limit, covering $0 < z < 1.2$, with about $584\,000$ galaxies in the W1 and W3 fields. Moreover, the galaxy selection in PAUS is genuinely magnitude limited. 
As we show in Section~\ref{sec:basic_results}, the requirements placed on the shape or features in a galaxy spectral energy distribution in order to measure a photometric redshift are less demanding than those needed to successfully extract a spectroscopic redshift. 
There is no requirement on finding spectral features to measure a redshift with a high degree of certainty, so there is no bias against objects with weak spectral breaks or emission/absorption lines. As part of their study of spectral features in PAUS galaxies, \cite{Renard:2022} looked at the evolution of galaxy colour for a sample matched to the VIPERS survey mentioned above. 

The redshift range covered by PAUS galaxies corresponds to a look back time of around 8 billion years or about two-thirds of cosmic history. 
Over this period a dramatic change took place in the global star formation rate (SFR) density  \citep{madau_dickinson_14}. 
The present-day SFR density is around one-tenth of the value at the peak, which occurred just above $z = 1$. 
Hierarchical models of galaxy formation have traditionally struggled to reproduce a drop in the global star formation activity of the same size (e.g. \citealt{Baugh:2005}, \citealt{Lagos2018}). 
The inference of the global SFR from observations is fraught with difficulties, such as accounting for the attenuation of starlight by dust, which is important at the short wavelengths that are most sensitive to recent star formation, and the `correction' for galaxies that are too faint to be observed. Instead, we take the more direct approach of considering observed galaxy colours rather than extracting model dependent quantities from observations. The $g-r$ colour is less affected by dust attenuation than the UV fluxes used to deduce SFRs. We will compare the predictions of galaxy formation models to the location and width of the red and blue clouds, and to the numbers of galaxies they contain.

Optical galaxy colours are sensitive to the star formation activity in galaxies and other intrinsic properties such as the metallicity and overall age of the composite stellar population and the galaxy stellar mass (e.g. \citealt{Daddi:2007, Taylor:2011, Conroy2013, Robotham:2020}). Galaxy colours are also correlated with morphology \citep{Strateva:2001}. Hence by measuring galaxy colours we can in principle constrain some of the physical processes that change the star formation history of a galaxy and the chemical evolution of its stars. The relative importance of gas cooling, and heating by supernovae and AGN is expected to change over the time interval accessible through the PAUS data.

The traditional way to analyse galaxy surveys, particularly ones that cover a substantial baseline in redshift, is to estimate rest-frame luminosities for galaxies. 
This involves correcting for band-shifting effects, which lead to filters in the observer frame sampling progressively shorter wavelengths in the rest frame of the galaxy with increasing redshift \citep{Hogg:2002,Kasparova:2021}. This correction depends on the shape of the galaxy's spectral energy distribution which depends on its star formation history, chemical evolution, stellar mass and dust content. Corrections may also be required for changes in the stellar populations over time, called evolutionary corrections in luminosity function studies \citep{Loveday:2015}. To accomplish this, the survey may be split into a set of disjoint redshift shells to measure the evolution of the luminosity function, however this results in removing many survey galaxies from the analysis. 

Here we take a simpler approach which uses all of the galaxies in a survey and tries to avoid any model dependent processing of the observations. 
We aim to compare the model predictions with actually observed quantities based on apparent magnitudes and redshift. 
In addition to basic statistics like the number counts and redshift distribution of galaxies, we also consider the evolution of the observed galaxy colours with redshift, exploiting the wide redshift baseline and homogeneous selection of PAUS.

To compare the evolution of observer frame colours with theoretical models it is necessary to include the sample selection and the band shifting effects in the model predictions. We do this by building a mock catalogue on an observer's past lightcone by implementing a semi-analytical model of galaxy formation into an N-body simulation \citep{Kitzbichler:2007,merson13}. This opens up a new set of tests of galaxy formation models: the overall galaxy number counts, the redshift distribution and the evolution of the observed colours; in the latter two cases, the statistics are measured for a specified magnitude limit. Hence, we extend the datasets typically used to calibrate galaxy formation models, such as the local luminosity function or stellar mass function, to include statistics that cover a range of redshifts and are relevant to ongoing surveys such as DESI \citep{DESI:2016,DESI:2022} and {\it Euclid} \citep{Euclid:2011}.  
We use the \galf galaxy formation model \citep{cole:2000,lacey:2016} implemented in the Planck Millennium N-body simulation \citep{Baugh:2019}. This extends the work of \cite{stothert18}, as the N-body simulation used here has superior resolution in mass and time. This allows us to include fainter galaxies in the mock catalogue and to make more accurate predictions for galaxy positions and luminosities. Also, since \cite{stothert18}, sufficient PAUS data has been collected to allow accurate measurements of the basic galaxy statistics to be made. 
A similar exercise was carried out by \cite{Bravo2020} who compared observed colours from a lightcone mock catalogues built using the \texttt{SHARK} semi-analytical model of \cite{Lagos:2019} to compare to the GAMA survey; here we extend this comparison to higher redshift. 

The remainder of the paper is laid out as follows: we first describe the theoretical framework used to build the PAUS mock in Sect.~\ref{sec:theo}, then we will present our main analysis and results in Sect.~\ref{sec:results}. In \S~\ref{sec:vary} we show how sensitive the model predictions are to the parameter choices. Finally, we present our conclusions in Sect.~\ref{sec:conclusions}.

\section{Theoretical model and observational dataset}
\label{sec:theo}

Here we describe the theoretical model, covering the galaxy formation model (\S~\ref{galaxy_formation}), the N-body simulation in which it is implemented (\S~\ref{sec:pmill}), the construction of the lightcone mock catalogue (\S~\ref{sec:lightcone}), before introducing the PAUS dataset in \S~\ref{sec:paus}.

\subsection{Galaxy formation model} 
\label{galaxy_formation}
We use the \galf semi-analytical model of galaxy formation \citep{cole:2000,bower:2006,lacey:2016}. The model follows the key physical processes that shape the formation and evolution of galaxies in the cold dark matter cosmology (for reviews of these processes and semi-analytical models see \citealt{baugh06} and \citealt{Benson:2010}). The model tracks the transfer of mass and metals between different reservoirs of baryons, predicting the chemical evolution of the gas that is available to form stars and the full star formation history of galaxies. When implemented in an N-body simulation, the semi-analytical model also provides predictions for the spatial distribution of galaxies \citep{Kauffman:99, Benson:00}.

The calibration of the model parameters is described in \cite{lacey:2016}, who provide a list of the model parameters in their table 1. Mostly local observational data is used in the calibration, which historically has been performed by hand in a `chi-by-eye' approach. \cite{elliott21} describe an automated and reproducible calibration that can perform an exhaustive search of a high-dimension parameter space.
Here we use the version of the model introduced by \cite{gonzalez14} (hereafter GP14), as recalibrated by \cite{Baugh:2019} following its implementation in the P-Millennium N-body simulation (which is described in Sect.~\ref{sec:pmill}). This recalibration required changes to the values of two parameters: the velocity that sets the mass loading of winds driven by supernova and the strength of AGN feedback (the parameter in this case effectively determines the halo mass at which AGN heating is able to stop gas cooling). We note that \cite{stothert18} used the \cite{gonzalez14} version of \texttt{GALFORM}; the model used here is essentially the same, with two small changes made to the parameter values as outlined above (see \citealt{Baugh:2019} for more details).

\subsection{The Planck Millennium N-body simulation}
\label{sec:pmill}
 
The Planck Millennium N-body simulation is part of the `Millennium' series of simulations of structure formation (\citealt{Springel:2005,Guo:2013}; see table~1 in \citealt{Baugh:2019} for a summary of the specifications of these runs and the cosmological parameters used). The Planck Millennium follows the evolution of the matter distribution in a volume of  $ 5.12 \,\times \,10^{8} \, {\rm Mpc}^{3}$, which is $1.43$ times larger, after taking into account the differences in the Hubble parameters assumed, than the simulation described by \cite{Guo:2013}, which was used by \cite{stothert18} to build an earlier mock catalogue for PAUS.

The Planck Millennium uses over 128 billion particles ($5040^{3}$) to represent the matter distribution, which is over an order of magnitude more than was used in the earlier Millennium runs. 
This, along with the simulation volume used, places the Planck Millennium at a resolution intermediate to that of the Millennium-I simulation of \cite{Springel:2005} (hereafter MSI) and the Millennium-II run described in \cite{Boylan-Kolchin:2009}.
The Planck Millennium has many more outputs than the MSI, with the halos and subhalos stored at 271 redshifts compared with the $\sim60$ outputs used in the MSI. Dark matter halo merger trees were constructed from the {\tt SUBFIND} subhalos \citep{Springel:2001} using the {\tt DHALOS} algorithm described in \cite{Jiang:2014} (see also \citealt{merson13}). Halos that contain at least 20 particles were retained, corresponding to a halo mass resolution limit of $2.12 \times 10^{9} \, h^{-1} {\rm M_{\odot}}$.

\subsection{Building a lightcone mock catalogue} 
\label{sec:lightcone}
The construction of a mock catalogue for a cosmological redshift survey can be accomplished in different ways, resulting in predictions with different accuracies, and which inform us to different extents about the physics behind galaxy formation. In principle, a simple approach would be to sample a population of galaxies randomly from an observed statistical distribution such as the luminosity function. However, this would lead to a catalogue with information limited to the property studied in the statistical distribution, ignoring any other properties and their relation with other observables. Moreover, the biggest limitation is that such a simplistic catalogue would not even be able to track the evolution of the galaxy population with redshift. 

To build a more realistic catalogue we need to track the evolution of the dark matter structures and populate the dark matter halos with galaxies at different epochs. Here, we make use of the Planck Millennium N-body simulation described in the previous section. To populate dark matter halos in the simulation with galaxies, we implemented the \galf semi-analytic model of galaxy formation on the merger histories of the dark matter halos extracted from the simulation. The combination of the Planck Millennium and \galf results in a physically motivated model which includes environmental effects associated with the merger histories of halos, and gives predictions for the spatial distribution of galaxies. \galf predicts the chemical evolution of the gas and stars in each galaxy, along with the size of the disc and bulge components and their star formation histories. The model outputs the mass-to-light ratios in a list of filters that are specified at run time. Along with the model for attenuation of stellar emission by dust described in \cite{cole:2000} (see also \citealt{lacey:2016}), this allows the model to predict the brightness or magnitude of the model galaxies in these bands.

\galf outputs the properties and positions of the galaxy population in the simulation box at a discrete set of redshift outputs. The \textit{lightcone} is built by interpolating galaxy magnitudes and positions between the values at these discrete redshifts, using the redshift at which the galaxy crosses the observer's lightcone. Thanks to the high time resolution of the Planck Millennium outputs, the reliability of the interpolation process described below is increased compared to that in earlier Millennium simulations.

To build the PAUS mock we follow the procedure described in \cite{merson13}.
We first place an observer at some position inside the simulation box, and choose a line-of-sight direction\footnote{It is good practice to choose a line of sight that does not coincide with one of the axes of the simulation box to maximise the distance (and hence time) between repetitions of the same structure.} for the mid-point of the survey, and a solid angle. Given the size of the simulation box, using this volume on its own we would only be able to probe redshifts out to $z \approx 0.19$. 
Hence, to cover the volume sampled by PAUS we need to replicate the simulation box in space using the periodic boundary conditions of the simulation. 
A galaxy crosses the past lightcone of the observer in between two of the simulation output redshifts or snapshots. The positions of the galaxy in the two snapshots are used to estimate its position at the lightcone crossing. \cite{merson13} applied different interpolation procedures for central and satellite galaxies. Central galaxies are assumed to be at the centre of mass of the host dark matter halo and hence track its motion between the snapshots. 
In this case, a simple linear interpolation is sufficient. Satellite galaxies, on the other hand, follow more complicated paths and can enter the observer's past lightcone either before or after their associated central. For this reason, a more sophisticated treatment is needed to compute the position of a satellite galaxy, taking into account its orbit around the central (see fig.~2 of \citealt{merson13}). Interpolating the galaxy positions in this way minimises artificial jumps in the correlation function measured from the lightcone.

Assigning properties to galaxies as they cross the observer's past lightcone using a simple interpolation between snapshots could lead to inaccuracies. 
The evolution of some properties, such as the SFR, is too complicated to be modelled by simple linear interpolation. Star formation can result from stochastic events, such as galaxy mergers and mass flows triggered by dynamically unstable discs, as well as smoother quiescent star formation in the galactic disc. 
For this reason, we follow \cite{merson13} and simply retain the galaxy properties from the higher redshift snapshot just above the redshift of lightcone crossing (as suggested by \citealt{Kitzbichler:2007}). Given the higher frequency of simulation outputs in the Planck Millennium run, the errors associated with this treatment are smaller than in previous Millennium simulations.

The one exception to this is the magnitude of the galaxy in the pre-specified filters in the observer frame. 
The definition of the observer frame depends on redshift and so is slightly different at the two redshifts that straddle the lightcone crossing redshift. We perform a linear interpolation between these two versions of the observer frame magnitudes to compute the observed magnitude at the redshift of lightcone crossing. 
In addition to the band shifting of the observer frame, we need to use the luminosity distance that corresponds to the lightcone crossing redshift to compute the apparent magnitude of the galaxy in the mock. 
This approach does not take into account any change in the spectral energy distribution of the galaxy between the higher redshift snapshot and the lightcone crossing redshift. 
However, the resulting colour-redshift relation is smooth and contains no trace of the locations of the simulation snapshots, as shown in Fig.~\ref{fig:nb_zz} in Appendix~\ref{app:0}. 
Here we test the interpolation scheme further by estimating photometric redshifts for the mock galaxies (see Section 3.2) and by looking at the colour redshift relation defined using colours obtained from the PAUS narrow band filters (Appendix~\ref{app:0}).  

Using the methods set out above, we have built a mock catalogue for PAUS which covers approximately $100 \,\rm{deg}^2$, with a magnitude limit of $i_{\rm AB} = 24$.  We used P-Millennium snapshots in the redshift range $0<z<2$. For some applications, we impose a magnitude limit to the mock of $i_{\rm AB}=22.5$\footnote{When this project was started, the bulk of the available PAUS photometry was limited to $i_{\rm AB}=22.5$. Since then deeper imaging has been processed and the majority of the catalogue is now limited to $i_{\rm AB} =23$.}. Some of the predictions we present include a simple model for errors in the photometry of \galf galaxies, which is set out in Appendix~\ref{app:a}. In this case, the magnitude limit is imposed after applying the perturbations to the raw magnitudes to account for the photometric errors. 

\begin{figure*}
\centering
    \subfloat{{\includegraphics[width=0.4\textwidth]{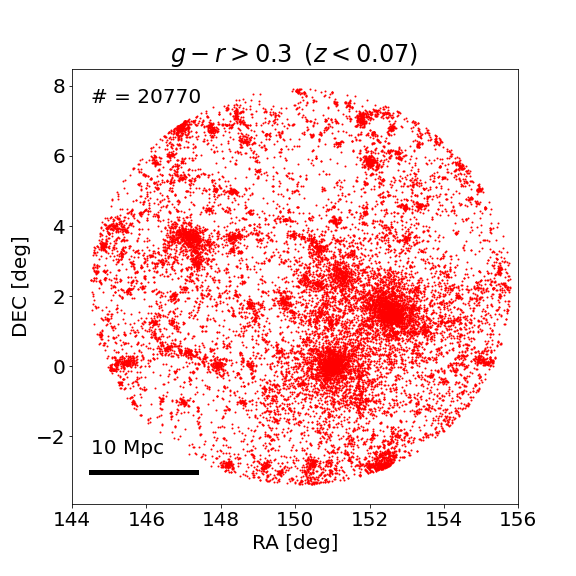} }}%
    \subfloat{{\includegraphics[width=0.4\textwidth]{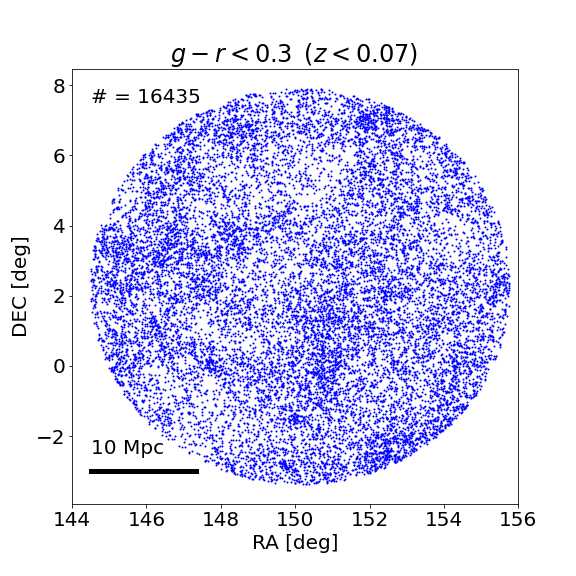} }}%
    
    \subfloat{{\includegraphics[width=0.4\textwidth]{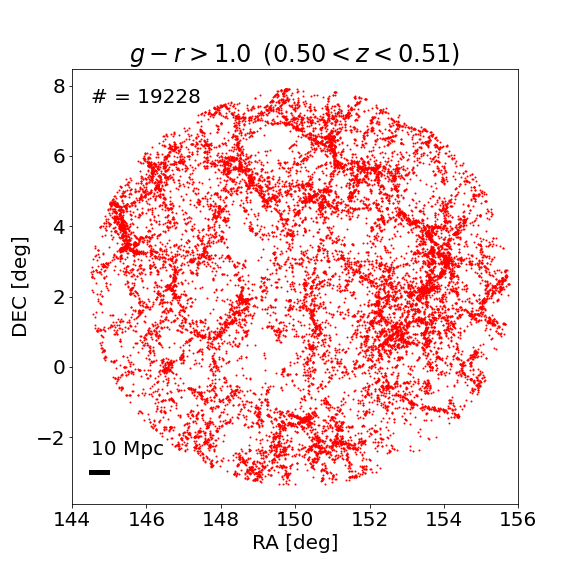} }}%
    \subfloat{{\includegraphics[width=0.4\textwidth]{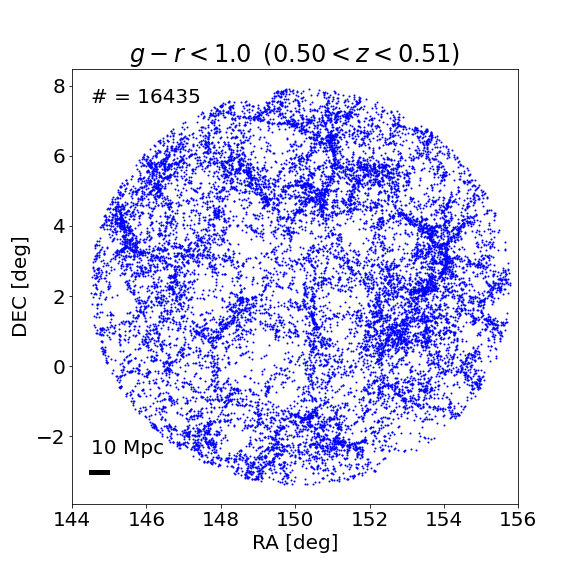} }}%
    
    \subfloat{{\includegraphics[width=0.4\textwidth]{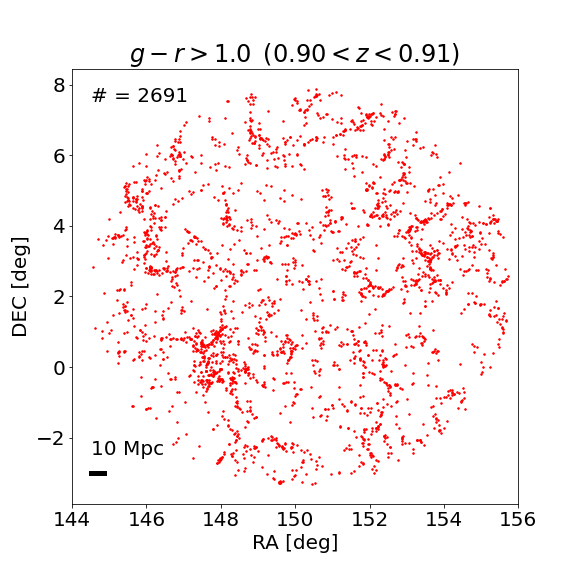} }}%
    \subfloat{{\includegraphics[width=0.4\textwidth]{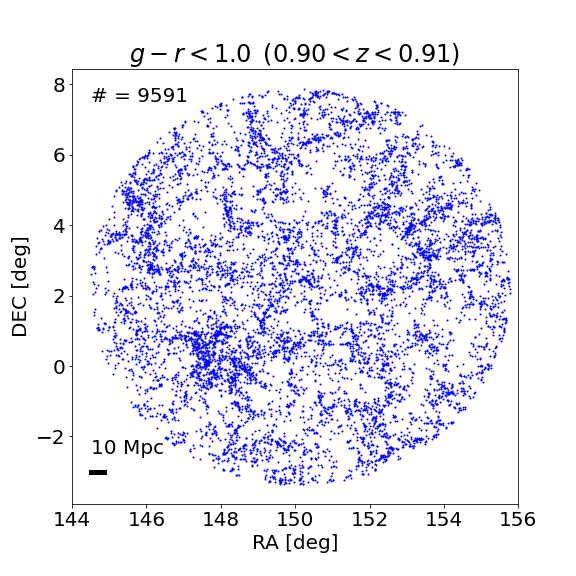} }}%
    
    \caption{Projected angular positions of galaxies in the lightcone mock catalogue (similar to right ascension and declination in degrees) in three different redshift intervals (as labelled),  separated into red (left column) and blue galaxies (right column) according to their observed $g-r$ colour (see Fig.~\ref{fig:colour_redshift_lightcone}). The lightcone covers approximately 100 $\rm{deg}^2$ and is magnitude limited to $i_{\rm{AB}} = 22.5$.  The presence of two big clusters at low redshift (top panels) can affect the number counts. For reference, the thick black bar in each panel indicates a scale of $10$ Mpc. The number of galaxies plotted is given in the top left of each panel.}%
    \label{fig:ra_dec}%
\end{figure*} 

\subsection{The PAU Survey}
\label{sec:paus}
We test the \galf lightcone against the Physics of the Accelerating Universe Survey (PAUS). PAUS was carried out using PAUCam \citep{padilla19}, a camera that was mounted on the William Hershel Telescope (WHT) in La Palma, Spain. PAUS is a novel imaging survey, with the key feature being the $40$ narrow-band filters of width $130$\AA\, covering the wavelength range from $4500$\AA\, to $8500$\AA\,, spaced by $100$\AA\, (see fig. 1 in \citealt{Renard:2022}). 
The 40 PAUS narrow bands overlap the wavelength range covered by the CFHTLenS $g$, $r$ and $i$ broadband filters \citep{Erben:2013}, as shown in fig.~1 of \cite{stothert18} and \cite{Renard:2022}. 
The narrow bands are particularly important when estimating photometric redshifts. The precision that PAUS can achieve is intermediate between that which can typically be achieved with a handful of broadband filters and that obtained with spectroscopy in a large-scale structure survey, in which case the spectral resolution and exposure time are chosen to maximise the number of redshifts that can be measured. \cite{eriksen19} report an error of $\sigma_z = (z_{\rm{photo}}-z_{\rm{spec}})/(1+z_{\rm{spec}}) \sim 0.0037$ when selecting the `best' $50$ per cent of the PAUS photometric redshifts in the COSMOS field limited at $i_{\rm AB}=22.5$. 
PAUS observations are available for the CFHTLenS wide fields: W1, W3 and W4, and the W2 field which corresponds to the Kilo Degree Survey (KiDS) \citep{Kuijken:2019}. For this study, we have decided to use the largest fields in PAUS which are W1, covering $13.71$ deg$^2$ and W3 covering $24.27$ deg$^2$ (giving a total of $37.98$ deg$^2$). We use photometric redshifts estimated using the BCNZ2 code  following the approach taken by \cite{eriksen19}.
We note that improved estimates of the photometric redshifts in PAUS have also been produced in a series of papers \citep{Eriksen2020,alarcon21,cabayol21,david23}.

\section{Results}
\label{sec:results}

We first describe some basic properties of the lightcone mock, such as its visual appearance, number counts and redshift distribution (\S~\ref{sec:basic_results}), before describing the estimation of photometric redshifts for the mock galaxies, using a simple, approximate model for flux errors (\S~\ref{sec:photoz}) and then comparing the evolution of the observed colours with PAUS (\S~\ref{sec:colour_redshift}). Finally, we assess the sensitivity of galaxy colours to the model parameters (\S~\ref{sec:vary}).

\subsection{Basic results: number counts and redshift distribution} 
\label{sec:basic_results}

In this section, we discuss the basic predictions of the simulated lightcone to show that they can reproduce the trends observed in the PAUS observations. One important feature of the lightcone is its magnitude limit cut.
For some purposes, the magnitude limit of $i_{\rm AB}=22.5$ is imposed on the magnitudes of mock galaxies without photometric errors. In other cases, the mock galaxy magnitudes are perturbed as described in Section~\ref{sec:photoz} and Appendix~\ref{app:a} and the magnitude limit is applied to a deeper catalogue to investigate the impact of photometric errors. The narrow band photometry has been computed using the transmission curves estimated by \cite{casas16} and \cite{padilla19} for the PAUCam optical system and the broadband photometry has been computed from the transmission curves used in the CFHTLenS \citep{Erben:2013}.

The distribution of the mock galaxies on the sky for three representative redshift bins is shown in Fig.~\ref{fig:ra_dec}, where we have split the galaxies into red and blue populations according to the observed $g-r$ colour (see Eq.~\ref{eq:redblue} and the associated discussion in Section~\ref{sec:colour_redshift}). 
The spatial scale in these images is indicated by the bar which shows a scale of $10$ Mpc, and allows us to compare the size of the structures at different redshifts. 
As shown in previous studies (e.g. \citealt{zehavi11} using Sloan Digital Sky Survey galaxies), red galaxies tend to cluster more strongly than blue ones. 
This is driven by environmental effects, such as the quenching of gas cooling and star formation when galaxies fall into the potential well of a more massive host dark matter halo (for example due to ram pressure stripping or other similar phenomena related to the removal of gas from galaxies due to gravity or tidal interactions). 
In the first row of Fig.~\ref{fig:ra_dec} ($0<z<0.07$), this effect is clearly visible with structures traced out by red galaxies being sharply defined compared to the more `diffuse' distribution of blue galaxies seen in the right panel. 
In the middle row of Fig.~\ref{fig:ra_dec} ($0.50<z<0.51$) as we zoom out, a larger region of the cosmic web is visible. 
The difference in the contrast of the structures seen with red or blue galaxies is now less pronounced, but still present, with the structures traced by blue galaxies appearing somewhat less sharp than those mapped by the red galaxies. 
In the bottom row of Fig.~\ref{fig:ra_dec}, which shows the redshift slice $0.90<z<0.91$, we can see that although the total number of galaxies is lower than it is in the other lower redshift bins, the relative numbers of red and blue galaxies are reversed (i.e. we now have more blue galaxies than red ones), due to the general uplift in star formation activity with increasing redshift. 

\begin{figure}
\centering
    {\includegraphics[width=0.5\textwidth]{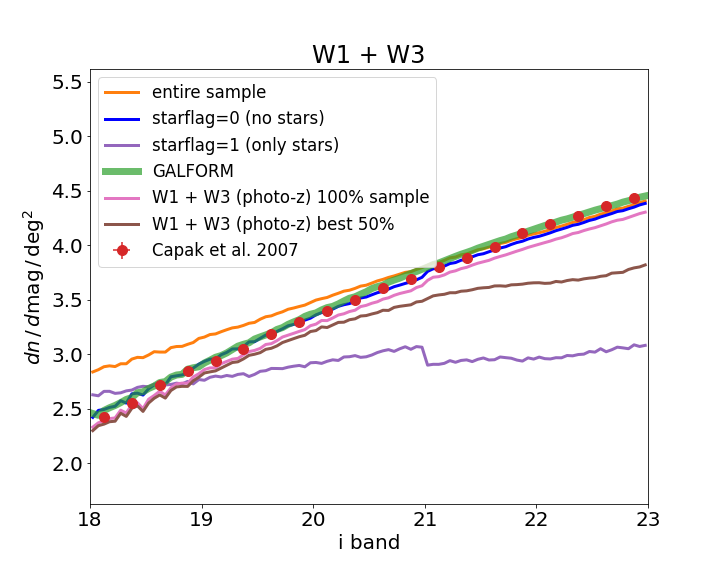}}%
\caption{Number counts in the $i$-band predicted from the \galf mock (thick green line) compared with the number counts from \protect\cite{capak07} (red points) and the PAUS data in the W1 and W3 fields for different selections: full photometric sample (orange line), this includes all objects that have been observed in the narrow-band (NB) filter NB455 (this means that they might not have a redshift estimate), objects with \texttt{star$\_$flag$=0$} (blue line) which are those that has been classified as galaxies from a CFHTLenS star-galaxy separation algorithm, objects with \texttt{star$\_$flag$=1$} (violet line) which are those that has been classified as stars, total photo-z sample (pink line), which are the galaxies that have a PAUS redshift estimate (they need to be observed in a large fraction of the NB filters) and $50$ per cent of the best quality redshift sample (brown line) according to the quality flag $Q_z$ as described in \protect\cite{eriksen19}. }%
    \label{fig:num_counts}%
\end{figure}

Now that we have gained a visual impression of the galaxies in the lightcone, and have seen how different colour populations trace out structures, we are ready to perform more quantitative analyses. The first simple characteristic measure of an optically selected galaxy sample is the number counts as a function of magnitude. 
We plot the $i$-band number counts in Fig.~\ref{fig:num_counts}.
The blue line in Fig.~\ref{fig:num_counts} represents an estimate of the observed galaxy number counts for PAUS in the W1 and W3 fields (which cover, respectively, areas of 13.71 deg$^2$ and 24.27 deg$^2$, giving a total of 37.98 deg$^2$). 
This is the area covered by the PAUS observations with at least one measurement in the narrow band filter at $455$ nm. This results in a more complete sample than the PAUS photo-z catalogue, because in order to measure a photometric redshift, there is a requirement for the galaxy to be imaged in at least 30 out of 40 narrow band filters (as well as the 5 CFHTLenS broadbands from the parent catalogue). This target is not always met for the PAUCam imaging \citep{padilla19}.  
We also include the number counts of the subsample of galaxies with photometric redshifts (pink line). The photo-z catalogue covers areas of $9.73$ deg$^2$ and $20.37$ deg$^2$ in W1 and W3 respectively, giving a total of $30.10$ deg$^2$ which is $79$ per cent of the photometric sample area. 
An important thing to note is that the shape of the number counts is the same for the photometric (blue line) and the photometric redshift (pink line) catalogues, which means that we expect their statistical properties to be similar, modulo a simple \textit{sampling factor} (the median ratio between the sample with photometric redshifts and the full photometric sample number counts) that we estimate to be about $0.897$. 
It is common practice in photometric redshift studies to apply cuts on the quality of the redshift estimates to define a new subsample of the catalogue for a particular analysis. The number counts for the best $50$ per cent of the photo-z sample are shown by the brown line in Fig.~\ref{fig:num_counts}. In this case, the shape of the number counts starts to depart from that of the photometric sample for magnitudes fainter than $i_{\rm{AB}}\sim 20$. This occurs because the fraction of objects with poorer quality factors increases as fainter magnitudes are reached. This is an important factor to consider when performing statistical tests and the impact of this cut on galaxy colours will be considered later on.

\begin{figure*}
\centering
    \subfloat{{\includegraphics[width=0.5\textwidth]{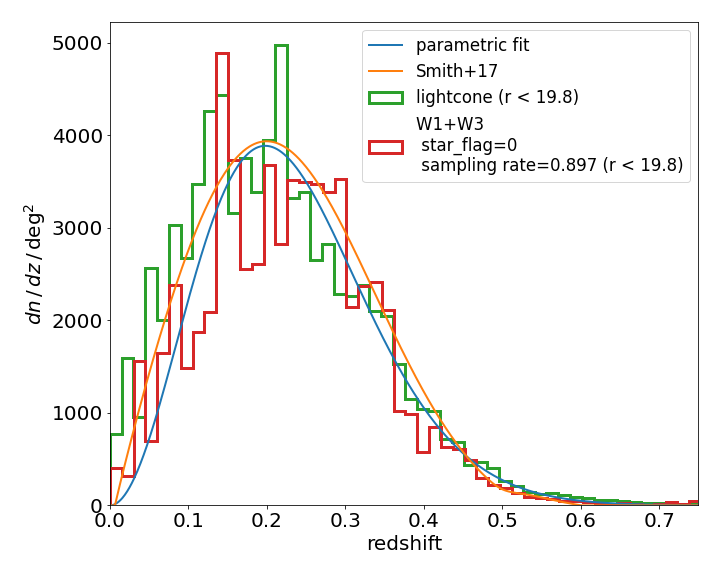} }}%
    \subfloat{{\includegraphics[width=0.5\textwidth]{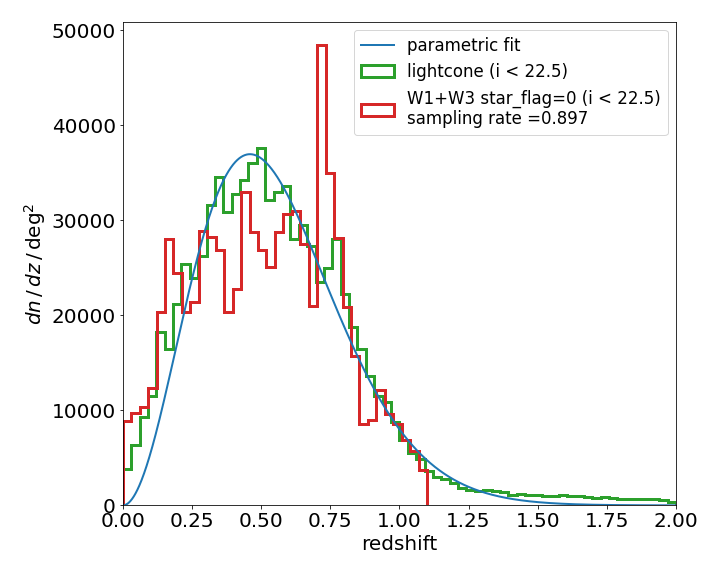} }}%
     \caption{The redshift distribution of galaxies brighter than $r_{\rm AB}=19.8$ (left) and $i_{\rm AB}=22.5$ (right). In both cases the red histograms show the measurements from the PAUS W1 and W3 fields combined, after imposing the \texttt{star}$\_$\texttt{flag}$=0$ cut to reject stars. The amplitude of the red histograms has been enhanced by dividing by the sampling rate factor stated in the legend, to take into account the fact that the photometric redshift catalogue is missing the fraction of objects that have less than 30 narrow band measurements. The green histograms show the lightcone redshift distributions, using the exact redshifts (i.e. the cosmological redshift plus the contribution of peculiar velocities predicted by the model) rather than the photometric redshifts that are discussed in Sec.~\protect\ref{sec:photoz}. The blue curves show a simple parametric fit to the green histograms (see text). The orange curve in the left panel shows a fit to the redshift distribution measured from the GAMA survey from \protect\cite{smith17}.}
    \label{fig:n_of_z}
\end{figure*}

The blue curve in Fig.~\ref{fig:num_counts} is the best estimate of the {\it galaxy} number counts, after applying a simple cut to remove stars from the photometric catalogue. The raw uncorrected counts of all objects in the PAUS photometric catalogue are shown by the orange curve. The property \texttt{star$\_$flag}, defined in the CFHTLenS catalogue, is used to remove stars. Objects with \texttt{star$\_$flag}$ = 1$, which are deemed to be stars, are shown by the purple curve. Note that there is a change in the methodology used to assign the \texttt{star$\_$flag} value at $i_{\rm AB}=21$. At brighter magnitudes than this, the size of the image is compared to the size of the point spread function, with unresolved objects being classified as stars. At fainter magnitudes, an object has to be unresolved {\it and} a good fit to a stellar template to be labelled as a star \citep{Erben:2013}. After removing stars in this way, the galaxy counts (blue curve), agree well with a previous estimate from the smaller COSMOS field $\sim 4\, \rm{deg}^2$) by  \cite{capak07} (red points; these counts extend to fainter magnitudes than shown in the plot). 
The number counts predicted by {\tt GALFORM}, measured from the lightcone, are shown by the green thick line. These agree remarkably well both with the COSMOS and PAUS measurements, particularly in view of the fact that mainly local observations were used to calibrate the model.

As a further test of the \texttt{GALFORM} predictions for galaxy number counts, we compare with the target density of galactic sources in the Dark Energy Spectroscopic Instrument Bright Galaxy Survey (DESI BGS) input catalogue estimated by \cite{omar20,Omar:2021} (see also \citealt{DESIBGS:2023}). \cite{Omar:2021}  find an integrated surface density of sources to $r_{\rm AB}=19.5$ of $808 \,\rm{deg}^{-2}$. In the \texttt{GALFORM} mock we find $837\,\rm{deg}^{-2}$ to the same depth, which agrees with the DESI value to within $5$ per cent. For PAUS, combining the W1 and W3 fields, we obtain a surface density of $719\,\rm{deg}^{-2}$, which is about 10 per cent lower than the DESI BGS value. However, we note that the combined area of the W1 and W3 fields (for the photometric sample) is $37.98 \,\rm{deg}$, i.e. $400$ times smaller than the imaging data used to obtain the DESI BGS estimate. Therefore, the counts from the PAUS fields could be subject to sample variance.

After the number counts, the next statistic to consider that characterises the galaxy population is the redshift distribution, the number of galaxies per square degree as a function of redshift. We show the redshift distribution of galaxies to two flux limits in Fig.~\ref{fig:n_of_z}, $r_{\rm AB}=19.8$ in the left panel, the depth of the deepest fields in the GAMA survey \citep{Driver:2011} and the PAUS limit\footnote{For the PAUS sample used in this work, the limit of 22.5 is the common choice among several publications, but we stress that for the newest PAUS photometric productions a magnitude cut of $i<23$ is adopted.} of $i_{\rm AB}=22.5$ in the right panel, which is substantially deeper.

The distribution of photometric redshifts in the combined W1 and W3 PAUS fields is shown by the red histograms in the panels of Fig.~\ref{fig:n_of_z}. These distributions are obtained by imposing the respective flux limits used in each panel, along with a selection on a star-galaxy separation parameter to reduce the contamination by stars (i.e. only retaining objects with \texttt{star$\_$flag = 0}). The normalisation of the redshift distribution has been corrected for the offset between the number counts of objects in the photometric sample and the photo-z sample (this is the sampling factor described above).
The left panel of Fig.~\ref{fig:n_of_z} also shows a fit to the observed redshift distribution from the GAMA survey, made by \cite{smith17}.\footnote{The equation for the fit used by \cite{smith17} is: %
\begin{equation}
\scriptstyle
    n_{\rm{GAMA}}(z) = N_1 \, z^a \,\cdot\, \exp{[-b \, z^c]} + (  0.5\,N_2(\rm{sign}[z-0.35] + 1) \,\cdot\, \exp{[-d \, z^e]}) + f \nonumber
\end{equation}
with parameter values: %
 $N_{1} = 2.71 \times 10^{4}$, $N_{2} = 1.96\times 10^{2}$, $a = 9.22\times 10^{-1}$, $b = 1.92\times 10^{1}$, $c = 2.44$, $d = 1.08\times 10^{-8}$, $e = -2.77\times 10^{1}$, $f = -2.60\times 10^{2}$.}
This agrees well with the distribution of photometric redshifts from the W1 and W3 PAUS fields, which together correspond to about one-fifth of the total solid angle probed by GAMA. 
Note that in the right panel of Fig.~\ref{fig:n_of_z}, by construction the photometric redshift code does not return redshifts above $z=1.1$. 
It is also clear from this panel that there is a preference for photometric redshifts around $z \sim 0.75$, which is a systematic in the estimation that is being investigated by the PAUS team, rather than due to large-scale structure; the feature at $z \sim 0.15$ is due to large-scale structure (see figure 13 of \citealt{david23}). At low redshift the survey samples a smaller volume than at high redshift and the redshift distribution is more susceptible to fluctuations due to features like clusters.

The green histograms in Fig.~\ref{fig:n_of_z} show the corresponding redshift distributions predicted using the \galf lightcone. A simple fit to the lightcone redshift distribution is given by $n(z) = A \, z^2 \, \exp{[-(z/z_{\rm c})^\alpha]}$ \citep{Baugh:1993}. We find the best fitting parameters to be $A = 321\,428$, $z_{\rm c} = 0.18$, and $\alpha = 1.7$ for the $r_{\rm AB}=19.8$ magnitude limited $n(z)$ (left panel). While for the $i_{\rm AB}=22.5$ magnitude limited $n(z)$, the best fit is given by $A = 610\,000$, $z_{\rm c} = 0.4$ and $\alpha = 1.6$. The predicted redshift distributions agree well with the observed ones for both magnitude limits shown in Fig.~\ref{fig:n_of_z}.

\begin{figure*}
   \centering
     \subfloat{\includegraphics[width=9cm]{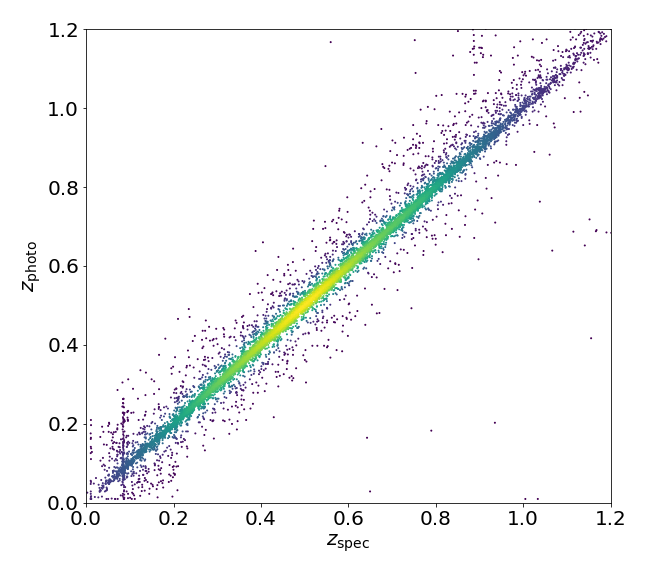}}%
     \subfloat{\includegraphics[width=9cm]{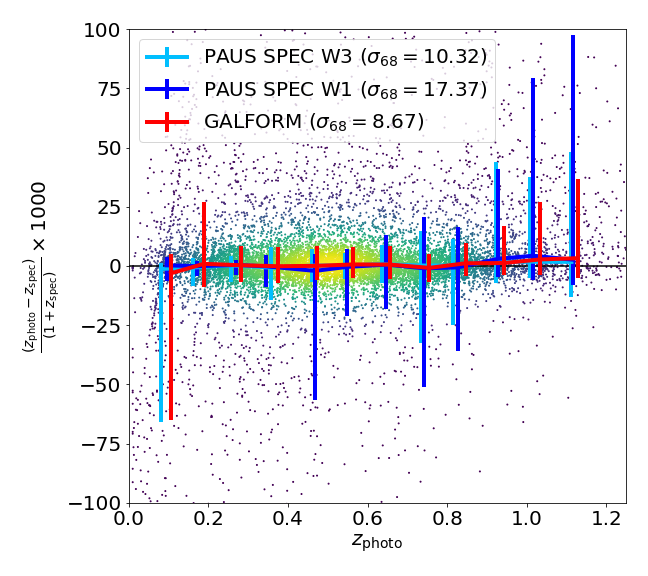}}%
\caption{Left: relation between the lightcone redshifts ($z_{\rm{spec}}$) and the photometric redshifts ($z_{\rm{photo}}$) obtained using the BCNz2 photo-z pipeline \citep{eriksen19}. 
The photometric redshifts are the results of running BCNz2 on the broad-band and the narrow-band filters with errors modeled using the prescription described in Appendix~\ref{app:a}. Right: relative error on the redshift estimated as the difference between photometric redshift and spectroscopic (lightcone) redshifts. 
The red line shows the median error in bins of redshifts for the \galf mock, with error bars indicating the 16$^{\rm{th}}$ to 84$^{\rm{th}}$ percentile range. 
The blue and light blue lines show the same quantity for a subsample of PAUS W1 and W3 respectively, matched with spectroscopic measurements from other overlapping surveys (for details, see \protect\citealt{david23}). 
The scale on the $y$-axis and the values the centralised $\sigma_{68}$ values quoted in the legend have been multiplied by 1000, to facilitate a comparison with the plots in \protect\cite{eriksen19}. 
In both panels, the points are from the mock and are coloured according to the density of points per pixel going from violet (low density) to yellow (high density).}
    \label{fig:zb_zspec}
\end{figure*}

\subsection{Estimating photometric redshifts for the mock}
\label{sec:photoz}

One of the important aims of this work is to quantify how the observed colour distribution of galaxies evolves with redshift (see Section~\ref{sec:colour_redshift}).
To isolate physical trends from those introduced by observational errors, we need to estimate photometric redshifts for the model galaxies. 
To do this, we need to model the observational errors in the photometry of the mock galaxies. 
We perturb the fluxes of the model galaxies to mimic the errors expected for the detection of a point source, given the magnitude limit of the PAUS observations in each band (see Table~\ref{tab:NBmaglim} in Appendix~\ref{app:a}; this appendix also discusses why we treat the galaxies as point sources). 
The errors are assumed to be Gaussian distributed in magnitude with a variance which is set by the signal-to-noise ratio at the magnitude limit in a particular band, using the formalism set out in \cite{errors:2020} (see Appendix~\ref{app:a} for more details). 
The broad band (BB) flux limits are much deeper than those for the PAUS imaging (see \citealt{Erben:2013}). 
The PAUS NB magnitude limits are $5 \sigma$ limits for point sources (see \citealt{Serrano:2023} and Table~\ref{tab:NBmaglim}).

The flux errors are computed for a subset of galaxies (44\,700) from the mock catalogue limited to $i_{\rm AB}=24$, which is a much deeper sample than the one we aim to analyse. This sample is then cut back to $i_{\rm AB}=22.5$ once the magnitude errors have been applied, giving a final sample of 14\,100 galaxies. The  BCNz2 algorithm \citep{eriksen19} is run on the perturbed model fluxes to estimate photometric redshifts. We then compare the scatter and fraction of outliers in the resulting photometric redshifts with those found for the observed galaxies.

Fig.~\ref{fig:zb_zspec} shows the results of this exercise. The left panel shows the estimated photometric redshift, $z_{\rm photo}$, as a function of the true value, $z_{\rm{spec}}$, which is the redshift including the effects of peculiar motions taken from the lightcone. This is the equivalent of a spectroscopic redshift but with no measurement error. We quantify the scatter in the photometric redshifts in a similar way to \cite{eriksen19}, using a centralised estimate, $\sigma_{68}$, defined as:
\begin{equation}
    \label{eq:sigma_68}
    \sigma_{68} = \frac{1}{2} \left( Q_{84} -Q_ {16} \right), 
\end{equation}
where $Q_{84}$ and $Q_{16}$ are the $84^{\rm th}$ and the $16^{\rm th}$ percentiles, respectively, of the distribution of the photometric redshift relative errors: $|z_{\rm photo} - z_{\rm spec}| / (1 + z_{\rm spec})$. 
This last quantity is plotted as a function of the estimated photo-z in the right panel of Fig.~\ref{fig:zb_zspec}. Estimates of the $\sigma_{68}$ are reported in the key of the same figure. 
The scatter found for the mock shares qualitative features with those inferred from the observations, being of the same order of magnitude and showing trends such as increasing with redshift. 
The observations that we use in the right panel of Fig.~\ref{fig:zb_zspec}, and that we label as `PAUS SPEC' are a match of the PAUS field W1 and W3 with spectroscopic measurements from other surveys\footnote{We need $z_{\rm spec}$ in order to estimate the relative error. This is because we assume that the spectroscopic redshifts have negligible uncertainties compared to photometric redshifts.}.
Since these PAUS SPEC samples are not simple flux-limited catalogues, they have a bias towards brighter magnitudes as a result of maximizing the number of spectroscopic redshift matches. 
The scatter predicted in the photometric redshifts for the mocks is nevertheless in reasonable agreement with the observational estimate. 

The characteristics of the mock photometric redshifts are discussed further in Appendix~\ref{app:b}. 
In summary, the size of the scatter is comparable to that estimated for the observations. 
However, the fraction of outliers is somewhat lower in the mock than in PAUS. 
This is due in part to our treating all of the model galaxies as unresolved point sources; in practice, resolved galaxies will have larger photometric errors, which could lead to more photometric redshift outliers. 
Also, we do not include the contribution of emission lines to the NB flux. 
The improved emission line model implemented in \galf by \cite{Baugh:2022} will be used in a forthcoming test of photometric redshift codes. 

\begin{figure*}
   \centering
    \subfloat{\includegraphics[width=9cm]{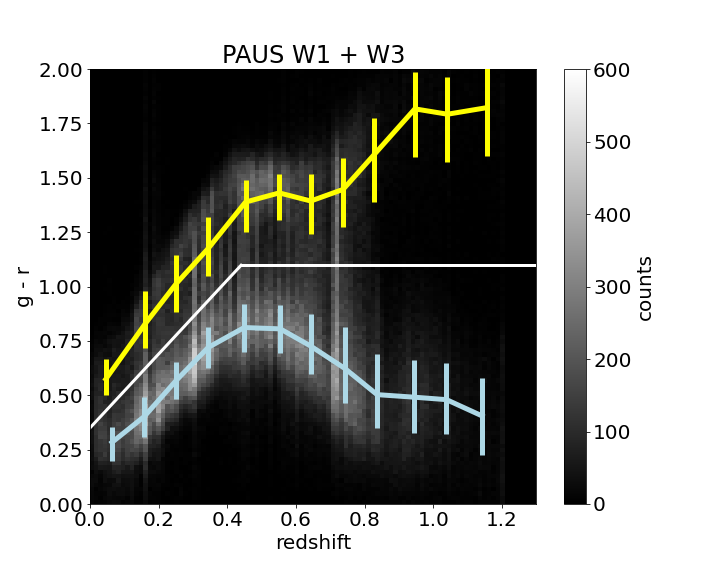}}
    \subfloat{\includegraphics[width=9cm]{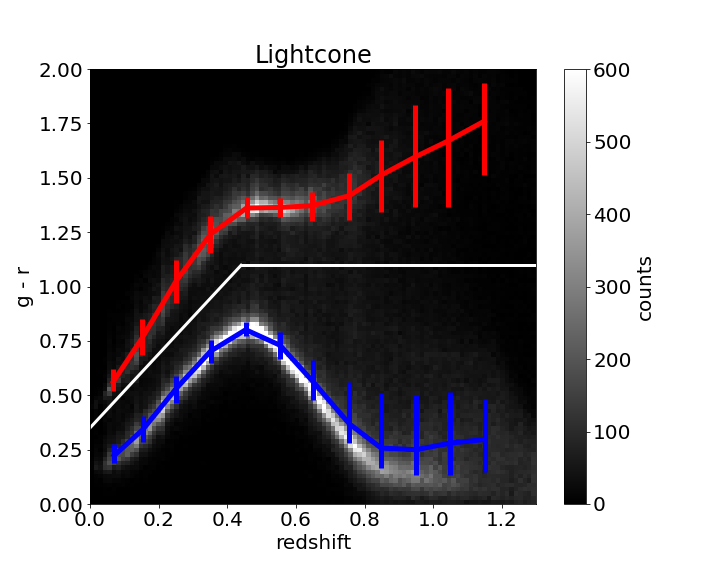}}
     \caption{2D histogram of galaxy counts in the observed $g-r$ colour vs redshift plane, for galaxies brighter than $i_{\rm AB}=22.5$. 
     The left panel shows galaxies from the combined PAUS W1 and W3 fields. The white line is used to separate red and blue galaxies (see text for equation). This is the same criteria used to separate red and blue galaxies in Fig.~\ref{fig:ra_dec}. Stars have been removed using the CFHTLenS property \texttt{star$\_$flag = 0}. The lines with bars show the median colour and 25-to-75 percentile range for the red and blue populations. The right panel shows the same plane for the model lightcone. As the model lightcone covers a roughly three times larger area than the observations, we have randomly sampled the model galaxies to match the total number of observed galaxies. To compare the two panels, we set the same colour bar; the most populated bins of the model lightcone are saturated with counts above the limit of 600 galaxies per bin. 
     }
     \label{fig:colour_redshift_lightcone}
\end{figure*}

Finally, it is reassuring that in Fig.~\ref{fig:zb_zspec} we can see no trace of any preferred values for the photometric redshifts recovered for the model galaxies. In particular, the redshifts of the original output snapshots in the N-body simulation are not apparent. 
This provides a validation of the methodology applied in order to compute the observer frame magnitudes in the model lightcone. 
Recall that the observer frame is defined at the simulation output redshifts on either side of the redshift at which the galaxy crosses the observer's past lightcone, and a linear interpolation is used to estimate the observer frame magnitudes in different bands at the lightcone crossing redshift \citep{merson13}. 
This point is investigated further in Appendix~\ref{app:0}.

The analysis in the subsequent subsections looks at the distribution of observed galaxy colours and their evolution with redshift. We will investigate the impact that errors in photometry and photometric redshift have on the \galf predictions.

\subsection{Evolution of galaxy colours} 
\label{sec:colour_redshift}
Here, we study the evolution of the observer frame $g-r$ colour with redshift. In an effort to keep the results from the observational data as model independent as possible, we use observer frame quantities to simplify the analysis, thereby avoiding the need to devise $k$-corrections to transform observed colours to the rest-frame.

\begin{figure}
   \centering
    \subfloat{\includegraphics[width=0.5\textwidth]{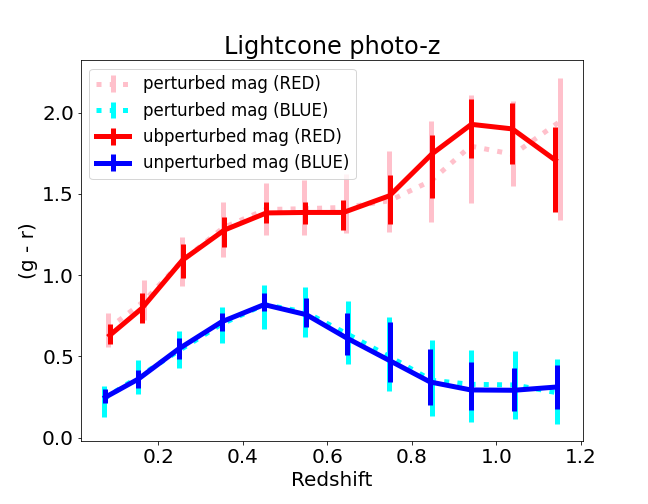}}
     \caption{Running medians for the observed $g-r$ colour vs redshift for \galf galaxies, comparing the case with (pink and cyan dotted lines) and without (red and blue lines) errors in the galaxy photometry and photometric redshift.}
     \label{fig:colour_redshift_comparison}
   \end{figure}

Fig.~\ref{fig:colour_redshift_lightcone} shows the distribution of galaxies with photometric redshifts in the observed $g-r$ colour $-$ redshift plane for the combined PAUS W1 and W3 fields (left) and the \texttt{GALFORM} model lightcone (right), in both cases to a magnitude limit of $i_{\rm AB}=22.5$. 
Note that in the \galf case in Fig.~\ref{fig:colour_redshift_lightcone} we are showing the galaxy colours without photometric errors and use the cosmological redshift (the effect of the inclusion of photometric errors and the use of the estimated photometric redshift is shown in Fig.~\ref{fig:colour_redshift_comparison} and discussed later in the text). 
Focusing on the left panel first which shows PAUS galaxies, the shading shows that there are two distinct populations of galaxies, the well known red sequence and blue cloud. Motivated by this, we place a dividing line to set the boundary between these populations:
\begin{eqnarray}
    \label{eq:redblue}
     g-r &=& 1.7 \, z + 0.35 \hspace{1cm} \text{ $z<0.44$,} \\ \nonumber
     g-r &=& 1.1 \quad \hspace{1.78cm} \text{ $z>0.44$}.
     \end{eqnarray}     
Blue galaxies lie below this line and red galaxies above it. 
Whilst there is a clear peak in the counts of galaxies in the red and blue clouds, there is a low count bridge of galaxies with intermediate colours connecting these two clouds. 
This is the so-called `green valley'. The minimum in the green valley is well defined and shifts to redder values of the observed $(g-r)$ colour with increasing redshift, up to $z \sim 0.4$. Beyond this redshift, the position of the green valley does not change in colour. The shape of the valley becomes more `flat bottomed' at high redshift, with the blue and red peaks moving further apart. At the highest redshifts the red peak becomes more indistinct and is much weaker than the blue peak.
Having split the population into two using this line, we can compute the median colours of the sub-populations on either side of the dividing line, along with the respective inter-quartile ranges (shown by the coloured lines and bars). 
The uneven density variations along the redshift axis are due to large-scale structure in the W1 and W3 fields.
\begin{figure*}
   \centering
    \subfloat{\includegraphics[width=0.53\textwidth]{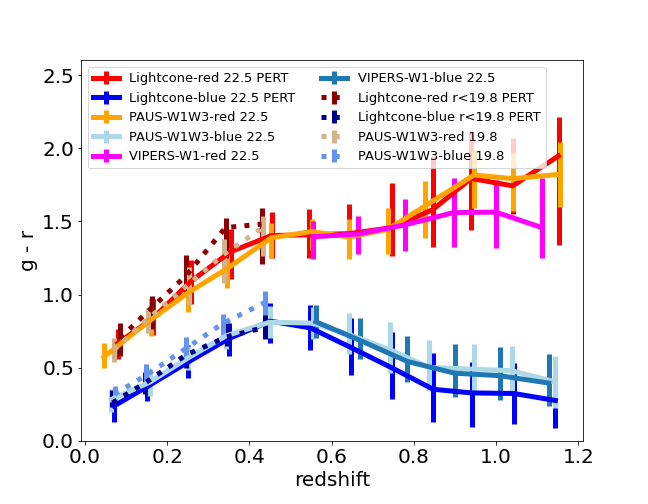}}
    \subfloat{\includegraphics[width=0.53\textwidth]{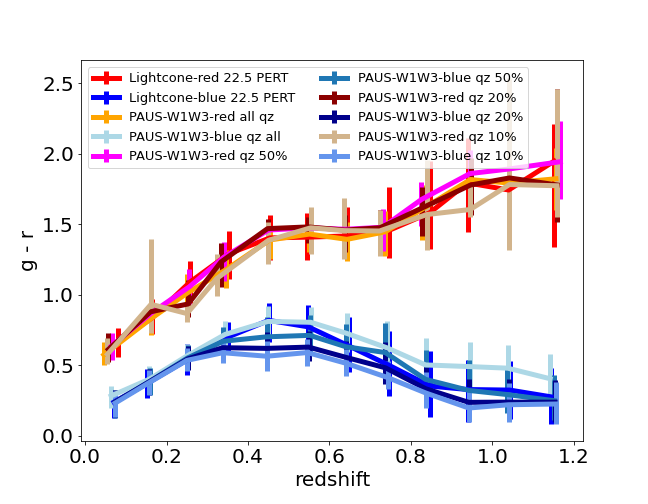}}
     \caption{Running medians for the observed $g-r$ colour vs redshift. In each case (lightcone, PAUS and VIPERS) red and blue galaxies have been split according to the white line in Fig.~\ref{fig:colour_redshift_lightcone} and the median has been computed in the two populations of galaxies separately. Left: the running median computed for different apparent magnitude limits. Right: the running median computed for different quality cuts, using the property $Q_z$ (see \citealt{eriksen19}) to identify the $50$, $20$ and $10$ per cent best quality redshifts in the sample.}
     \label{fig:colour_redshift_comparison2}
   \end{figure*}
\begin{figure*}
\centering
    {\includegraphics[width=1.\textwidth]{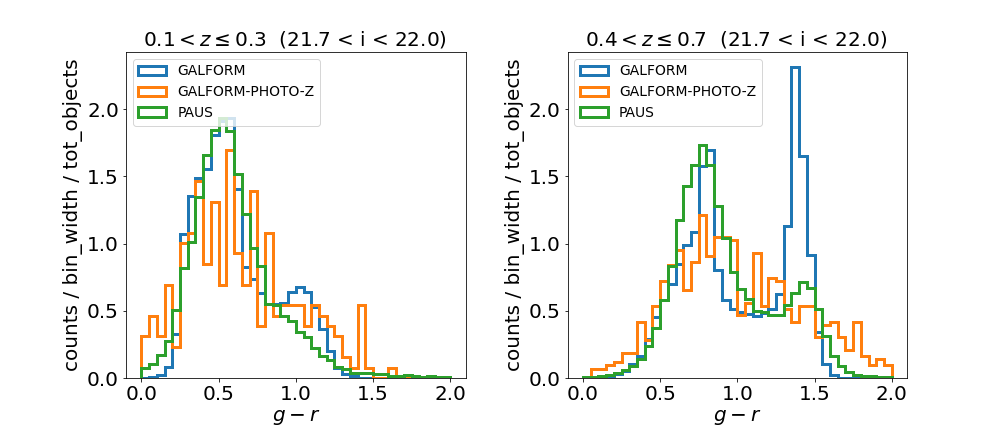} }%
    {\includegraphics[width=1.\textwidth]{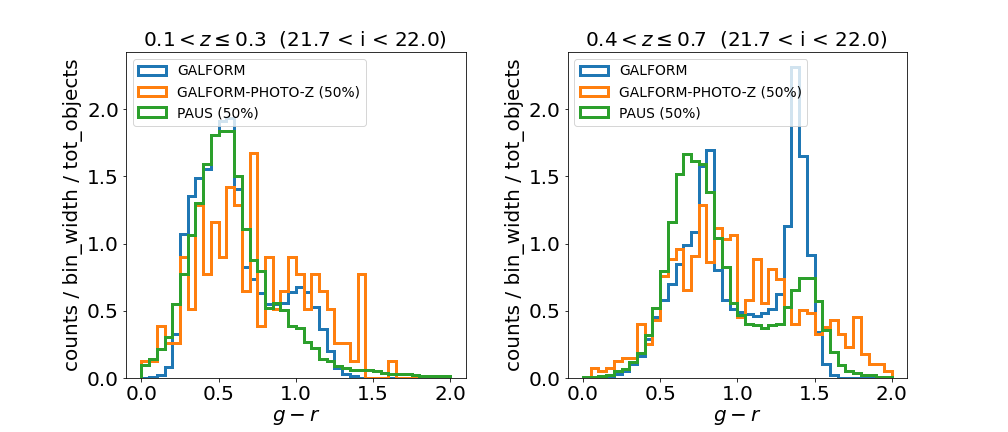} }%
\caption{Histograms of the observed $g-r$ colour for different redshift bins (different columns) in a selected narrow bin in the i-band ($21.7< i_{\rm{AB}} <22.0$), for the lightcone (unperturbed, blue histogram labelled as GALFORM and perturbed, orange histogram labelled as GALFORM-PHOTO-Z) and PAUS W1 + W3 fields (green histogram). The two rows show different cuts for the quality of the photometric redshifts (the unperturbed lightcone, blue histogram, is unchanged as it uses cosmological redshifts directly from the simulation). Specifically, the first row is for the full sample while the second row retains the best $50\%$ of objects according to the $Qz$ criteria. We note that for GALFORM-PHOTO-Z, the photometric redshifts are used to select the sample plotted.} 
    \label{fig:gr_histo}%
\end{figure*}

The observed $g-r$ colour evolves with redshift. 
There are two main physical contributions to the shape of the galaxy spectral energy distribution which affect this evolution: the attenuation of the starlight by dust and the shape of the stellar continuum. 
The latter effect depends on the amount of ongoing star formation and the age of the composite stellar population. 
In the rest frame, the effective wavelength of the $g$-band is $4792.9$\AA$\,$  and for $r$ it is $6212.1$\AA. 
The main spectral feature at these wavelengths, particularly once a modest redshift is applied to the source, is the 4000\AA \, break, a combination of various metal absorption lines over a range of several hundred Angstroms which are stronger in older stellar populations. 
PAUS images galaxies using narrow band filters that span the wavelength range from $4500$\AA\, to $8500$\AA\,. A wavelength of $4000$\AA \, in the rest-frame is sampled by the $g$ and the $r$ bands for redshifts in the range $0.16<z<0.36$.
The decline in the spectrum associated with the $4000$ \AA\, break actually starts around $4500$ \AA, close to the effective wavelength of the $g$-band. 
As redshift increases, the $g-$band in the observer frame samples progressively shorter wavelengths in the rest-frame, towards the $4000$\AA$\,$ break (see \citealt{Renard:2022} for a discussion of this spectral feature). 
The observed $g-r$ colour gets redder with increasing redshift, with the gradient being somewhat steeper for red galaxies (with deeper 4000\AA\, breaks). 
Note that star-forming galaxies display a modest reddening of the stellar continuum around 4000 \AA , albeit not as pronounced as in galaxies with older composite stellar populations. 
Hence the observer frame $g-r$ colour for star-forming galaxies in the blue cloud also gets redder with increasing redshift. 
At $z=0.3$, the observer frame $r$-band samples the rest-frame effective wavelength of the $g$-band at $z=0$, and the $g$ filter starts to move down to shorter wavelengths than the break. 
At higher redshifts than this, there is a divergence in the observer frame $g-r$ colours found for the red sequence and blue cloud, with both filters now sampling rest-frame wavelengths that are bluewards of the 4000\AA\, break. 

The right panel of Fig.~\ref{fig:colour_redshift_lightcone} shows the equivalent information for the model lightcone. As the model lightcone covers a much larger solid angle than the combined W1 and W3 fields, we have randomly sampled the model galaxies to match the total number of galaxies in the observed sample ($583\,992$ galaxies). To ensure that the random sampling does not affect the results we tested three different random seeds and observed no difference in the resulting colour redshift distribution. 
In principle, using the same number of objects allows us to use the same colour scale for the density shading for the observations and the model. 
However, as the colour bimodality is noticeably tighter in the model, the white bins in the right panel are all saturated as the counts reach around a thousand per pixel, and the colour shading peaks at 600 galaxies per pixel.
The larger solid angle of the model lightcone also means that large-scale structure has a smaller impact on the number of galaxies, so we see little evidence of any striping in redshift. 
The overall locus of galaxies in the red sequence and blue cloud in the model is similar to that seen in the observations, so we are able to use the same line to divide the model galaxies into red and blue subsamples.

To make a more quantitative comparison of the colour evolution between the observations and the model, we compute the median and interquartile range of the distribution of $g-r$ colour in narrow redshift bins, considering the blue and red populations separately. 
As we have already noticed by the relative tightness of the shaded regions in the colour-redshift plane in Fig.~\ref{fig:colour_redshift_lightcone}, the bimodality is stronger in the model colours compared with the observed ones. 
This is backed up by the narrower interquartile range of colours in the model compared with the observations. 
This behaviour of the model had already been noticed in previous comparisons \citep{gonzalez09,manzoni21}.

The predicted width of the red and blue populations is strongly affected by the addition of photometric errors (see Appendix~\ref{app:a} for a description of the errors applied), as shown by the inter-quartile ranges plotted in Fig.~\ref{fig:colour_redshift_comparison}. 
This figure shows the comparison in the running medians and percentiles for the colour - redshift relation when using the unperturbed colours $g-r$ predicted by the lightcone (red and blue lines) versus using the perturbed colours by adding the simulated errors (pink and cyan dotted line) as in Appendix~\ref{app:a}. 
The perturbed colours are plotted against the photometric redshift estimated by the photometric redshift code as in Section~\ref{sec:photoz}, while the unperturbed magnitudes are plotted against the cosmological redshift outputted by \galf.
This effect will be shown in plots of the colour distribution for different selections in redshift and apparent magnitude in the remaining of this section.  

We make further comparisons between the evolution of the observer frame colour distributions in the model lightcone and observations in Fig.~\ref{fig:colour_redshift_comparison2}, again including the effects of photometric errors in the model colours and using the estimated photometric redshifts. 
For clarity, we drop the density shading in this plot and show only the median colour and inter-quartile range for different selections. 
Note that the results for the model and the observations are plotted together in the same panel in this plot. 
The left panel of Fig.~\ref{fig:colour_redshift_comparison2} extends the standard colour - redshift comparison made at the PAUS depth of $i_{\rm AB}=22.5$ in two directions. 
First, we consider a brighter magnitude cut, $r_{\rm AB}=19.8$, which corresponds to the depth of the faintest fields in the GAMA survey. 
As expected, median colours can only now be plotted out to a lower redshift of $z=0.45$, as there are very few galaxies at higher redshifts. 
The median colours in the model are insensitive to this change in magnitude limit, though the observations suggest that both red and blue galaxies get redder with the brighter apparent magnitude cut. 
In the left panel of Fig.~\ref{fig:colour_redshift_comparison2} we also compare the model with an alternative sample of higher redshift galaxies, using the VIPERS spectroscopic sample \citep{scodeggio18},  which is limited to the same depth as PAUS $(i_{\rm{AB}}=22.5)$. 
Colour pre-selection is used to identify VIPERS targets, which limits this survey to redshifts $z\gtrsim0.5$ (see fig.~3 of \citealt{guzzo14} for the colour-colour selection used to select high redshift target galaxies). 
The high redshift tail of the colour - redshift relation agrees well between VIPERS and PAUS, suggesting that this result is not sensitive to errors in the estimated photometric redshifts and that the colour preselection in VIPERS is effective. 
This comparison shows the usefulness of the PAUS measurements which span a much wider redshift baseline than comparable spectroscopic surveys, which are either shallower and hence only cover the lower redshift half of the PAUS redshift range, as is the case with the GAMA survey, or which do not measure low redshift galaxies, as in the case of VIPERS.

The right panel of Fig.~\ref{fig:colour_redshift_comparison2} examines if the selection of higher quality photometric redshifts changes the appearance of the colour-redshift relation.
\cite{eriksen19} and \cite{alarcon21} show that the quality factor property can be used to define a subset of galaxies with fewer redshift mismatches or outliers and a smaller scatter in the estimated redshift than would be found in the full apparent magnitude limited sample. 
We want to rule out two effects: firstly that the distribution of quality factors might be different for red and blue galaxies due to a dependence of photometric redshift accuracy on galaxy colour, and secondly, that changing the fraction of outlier redshifts could alter the appearance of the colour - redshift relation. 
In the right panel of  Fig.~\ref{fig:colour_redshift_comparison2}, we plot the median colour for the entire sample, and for subsamples comprising the best $50$, $20$ and $10$ per cent of redshifts. Although the median colours agree within the $25$th - $75$th interquartile range, we note a slight shift in the blue cloud medians to bluer colours when restricting the sample to better quality redshifts. The colours measured for better quality photometric redshift samples seem to agree better with the lightcone predictions.

Finally, we dig deeper into the evolution of galaxy colours by considering galaxies selected to be in narrow ranges of apparent magnitude and redshift. 
In Fig.~\ref{fig:gr_histo}, we plot the distribution of the observed $g-r$ colours for both the \texttt{GALFORM} and the PAUS samples. 
We select a narrow apparent magnitude bin, $21.7<i<22.0$, to minimize the effect of having different galaxy populations\footnote{We would not need this requirement if using rest frame absolute magnitudes, as a specific luminosity would not vary with redshift, but that would lead to other problems such as using model-dependent k-corrections.} and study how this distribution change in two redshift bins: a  `low redshift' one spanning $0.1<z<0.3$ and a `high redshift' one covering $0.4<z<0.7$. 
As noted when commenting on Fig.~\ref{fig:colour_redshift_lightcone}, the bimodality of the colour distribution predicted in the \galf model, before the application of any errors in the galaxy photometry (blue histogram), is more pronounced than that seen in the observations (green histogram). 
This is quite clear in the high redshift bin. 
Including the simple model of photometric errors described in Appendix~\ref{app:a}, the bimodality in the \galf predictions that is prominent in the high redshift panels is greatly reduced (orange histogram). 
This brings the model into much better qualitative agreement with the observations. 
Reassuringly, the shape of the PAUS distribution does not change when selecting the best $50$ per cent of photometric redshifts using a cut on the quality parameter (bottom panels of Fig.~\ref{fig:gr_histo}). 
In the same way, the \galf predictions display similar behaviour when selecting the half of the sample with the best photometric redshifts. 

\begin{figure}
\centering
    {\includegraphics[width=0.48\textwidth]{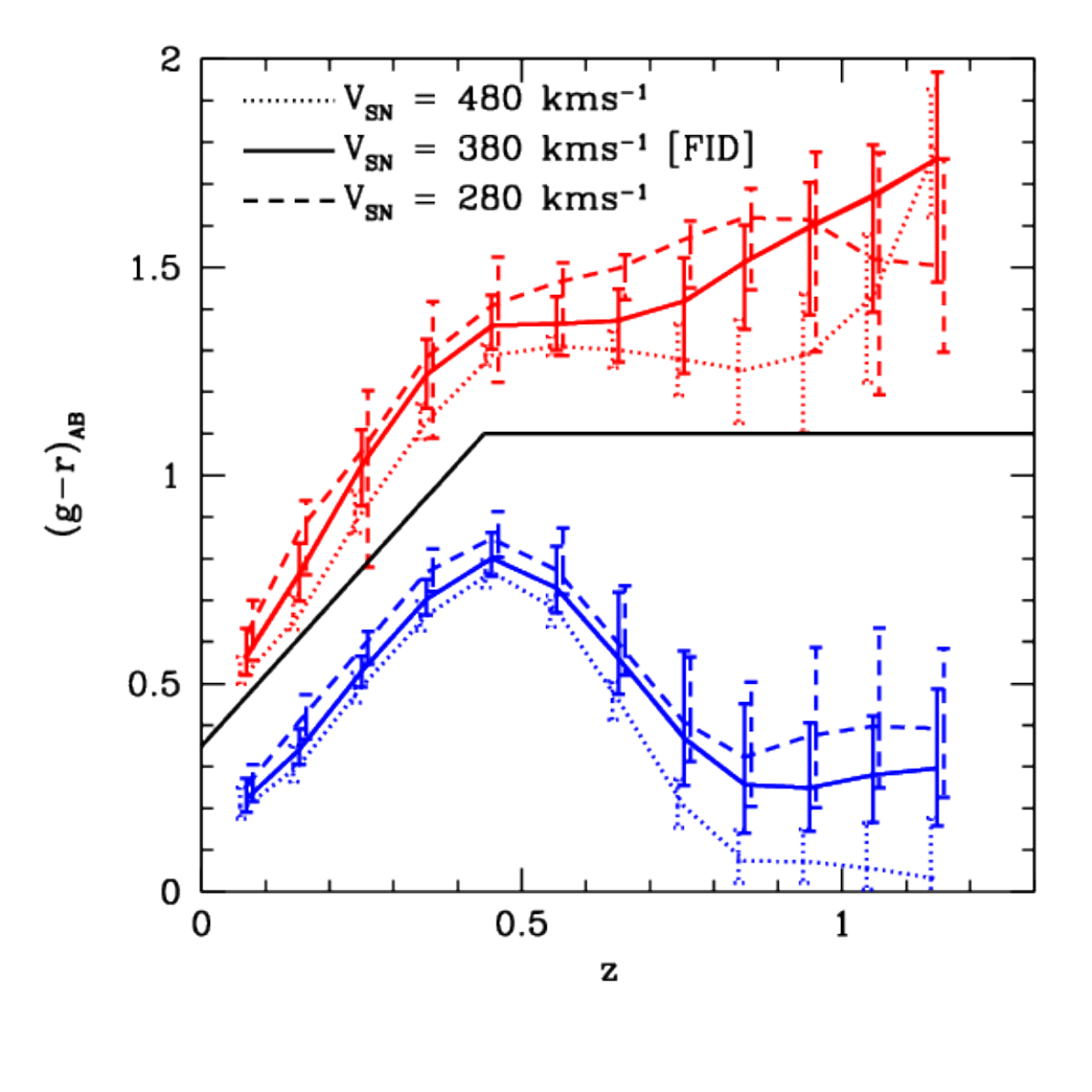} }%
    {\includegraphics[width=0.48\textwidth]{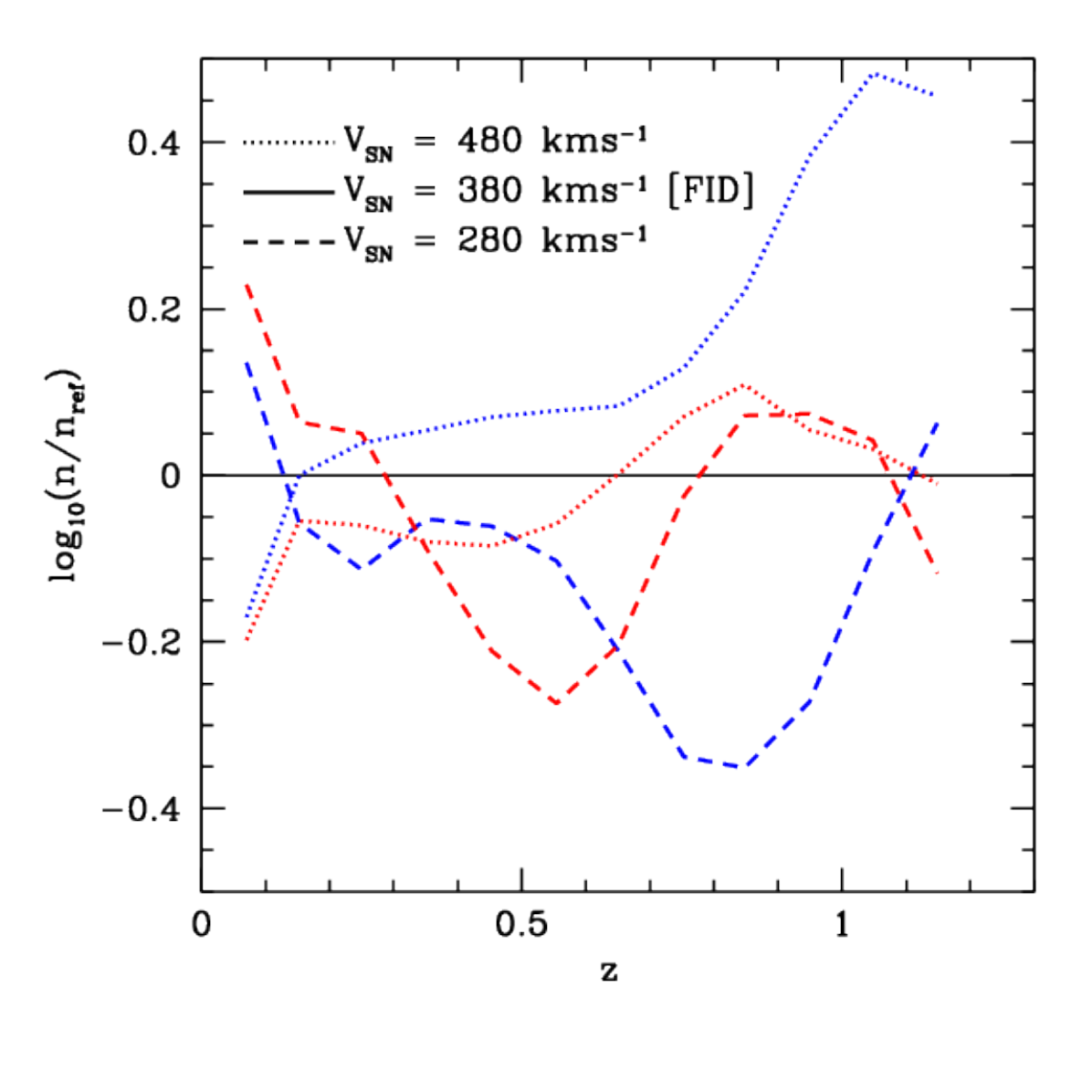} }%
\caption{The sensitivity of observed galaxy colours to the strength of SNe feedback, for samples limited to $i_{\rm AB} = 22.5$. In the default model, the scale velocity used in the mass loading of the SNe-driven wind is $V_{\rm SN}= \,380 \, {\rm km s}^{-1}$. The colour redshift relation for this model is shown by the solid lines in the top panel. The bars indicate the 25-75 percentile spread of the colours for the red and blue galaxy populations separately (i.e. those which fall on either side of the black line). The variant models correspond to $V_{\rm SN}= \,280\, {\rm km s}^{-1}$ (dashed) and $V_{\rm SN}= \,480 \,{\rm km s}^{-1}$ (dotted). The interquartile range is also shown for the variant models in the corresponding line style. The upper panel shows that these changes result in a shift in the median colours of the red and blue populations. The lower panel shows the logarithm of the number of objects in a specific population (red or blue as per the line colour) in the variants (line style) normalised by the number of objects in the same population in the fiducial model. The $i$-band luminosity functions in the variants have been rescaled to match that in the fiducial model at $z=0$, which affects the sample of galaxies plotted, but not their colours. }
    \label{fig:gr_var}
\end{figure}

\subsection{Sensitivity of galaxy colours to model parameters}
\label{sec:vary}

In this section, we explore the sensitivity of the observed galaxy colours to variation of the \texttt{GALFORM} model parameters. 
In particular, we look to see if altering the value of a parameter modifies the number of objects in the red and blue populations, or indeed produces a shift in the median colours of these populations. 
We focus on a subset of the processes in the model for this exercise, as they are known to have a big effect on the intrinsic galaxy properties by altering the star formation activity and hence affecting their colours. These processes are: the strength of the supernova (SNe) driven winds, the timescale for SNe heated gas to be reincorporated into the hot halo, the efficiency of AGN suppression of gas cooling, and the timescale for quiescent star formation. 

When a calibrated galaxy formation model is run with a perturbed value for one of its parameters, this can result in a change in the predictions for the observations used to calibrate the model (see the plots illustrating the impact of changing a range of model parameters in \citealt{lacey:2016}). 
In principle, other model parameters might need to be adjusted to ensure that the variant model produces an acceptable match to the calibration data, for example, using the methodology introduced by \cite{elliott21}. Here, we instead rescale the model galaxy luminosities to force agreement with the $i$-band luminosity function at $z=0$ predicted by the fiducial model. We chose the $i$-band as this is the selection band for PAUS. The same rescaling is applied at all redshifts, and to all bands. 
Hence, the rescaling does not change the model predictions for observer frame colours, but does affect which galaxies are selected to be part of the $i$-band apparent magnitude limited sample. Note that although, as we shall see below, in some cases the shape and location of the red and blue peaks can change, we have checked that the line separating galaxies into red/blue populations works equally well in all models and Eqn.~\ref{eq:redblue} is retained throughout. 

\begin{table}
	\centering
	\caption{The parameter values explored in the variant models. The first column gives the parameter name. The third column gives the fiducial value of the parameter, whereas the second and fourth columns give the low and high values considered, respectively.}
	\label{tab:variants}
	\begin{tabular}{lccc} 
		\hline
       Parameter &    low       & fiducial    &  high \\
        name      &          &     & \\
		\hline
		$V\textsubscript{SN}$ (km s$^{-1}$)&  280 & 380  & 480\\
		$\alpha\textsubscript{reheat}$ &  1.00 & 1.26 & 1.50\\
  		$\nu\textsubscript{SF}$ &  0.20 & 0.50  & 1.70\\
            $\alpha\textsubscript{cool}$ &  0.50 & 0.72 & 0.90\\		
		\hline
	\end{tabular}
\end{table}

Four model parameters are changed in this exercise, one at a time, resulting in eight variant models. The parameter values are listed in Table~\ref{tab:variants}: (i) the pivot velocity that controls the mass loading of SNe driven winds, $V_{\rm SN}$ (Eqn 10 in \citealt{lacey:2016}), with higher values resulting in larger mass ejection rates from more massive halos  (ii) the timescale for gas heated by SNe to be reincorporated into the hot gas halo, which is inversely proportional to $\alpha_{\rm reheat}$ (Eqn 11 in \citealt{lacey:2016}\footnote{In Lacey et~al. $\alpha_{\rm{reheat}}$ was called $\alpha_{\rm{ret}}$.}), with larger values giving shorter reincorporation times (iii) the star formation efficiency factor, $\nu_{\rm SF}$, (Eqn.~7 of \citealt{lacey:2016}; the variants listed in Table~\ref{tab:variants} correspond to the full range suggested by observations of local star forming galaxies \citealt{Blitz:2006}), and (iv) the factor which determines the halo mass in which AGN heating starts to prevent the cooling of gas, $\alpha_{\rm cool}$, (Eqn.~12 in \citealt{lacey:2016}). 
\begin{figure}
\centering
    {\includegraphics[width=0.48\textwidth]{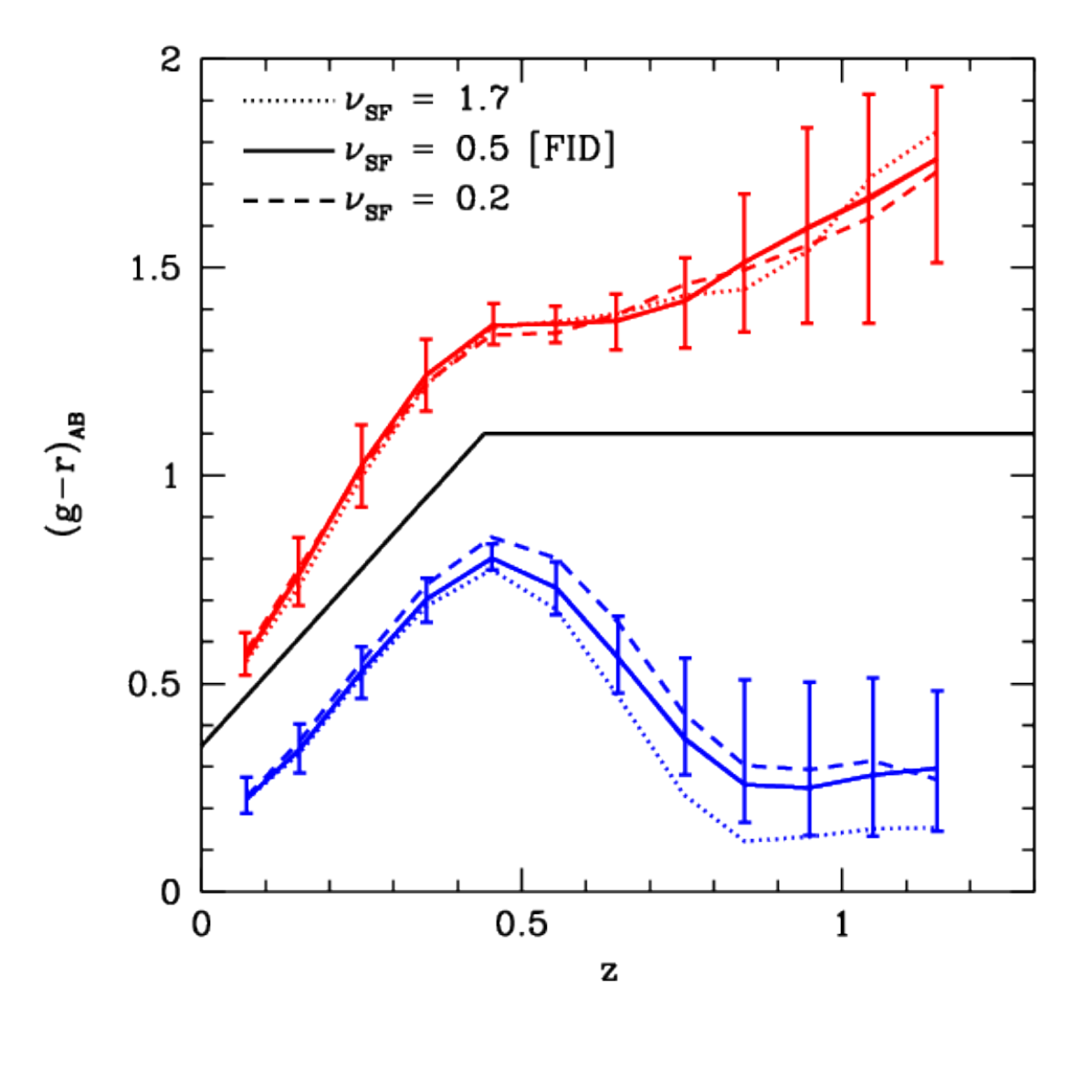} }%
    {\includegraphics[width=0.48\textwidth]{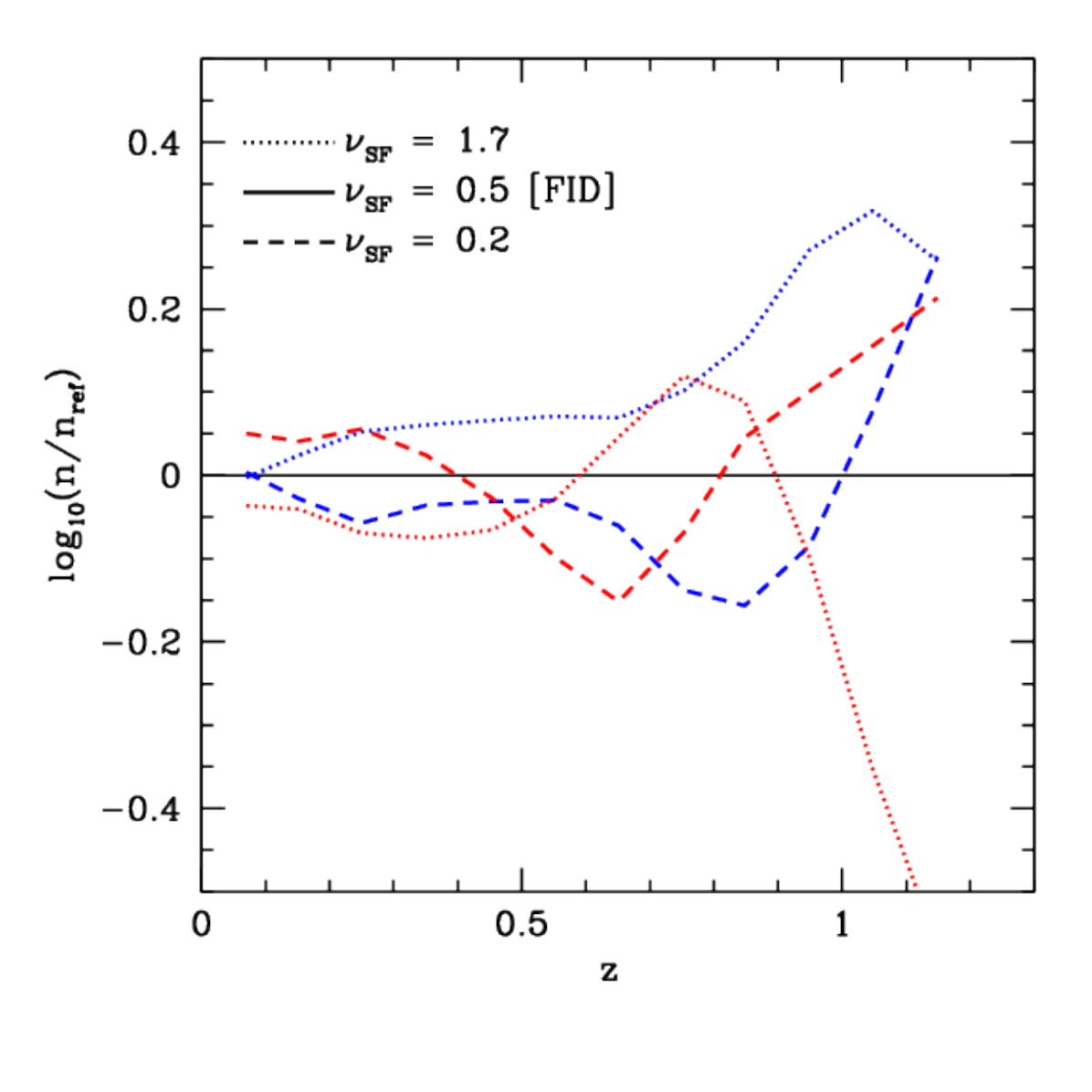} }%
\caption{The effect on galaxy colours of changing the star formation efficiency parameter, $\nu_{\rm SF}$. The top panel shows the median $g-r$ colour versus redshift, along with the 25-75 percentile range. The value in the fiducial model is $\nu_{\rm SF}=0.5$. The dashed line shows the predictions with $\nu_{\rm SF}=0.2$ and the dotted line shows $\nu_{\rm SF}=1.7$. This range for the parameter $\nu_{\rm SF}$ is inferred from observations (see \citealt{Blitz:2006}). With the same colour and line scheme as the upper panel, the bottom panel shows the log of the ratio between the number of galaxies in the variant model and the number of galaxies in the fiducial model for the desired population (red or blue). Note that in this and subsequent plots (Figs. 11 and 12), the interquartile colour ranges for the variant models are similar to those for the fiducial model and so are not shown.}%
    \label{fig:gr_var2}%
\end{figure}
\begin{figure}
\centering
    {\includegraphics[width=0.48\textwidth]{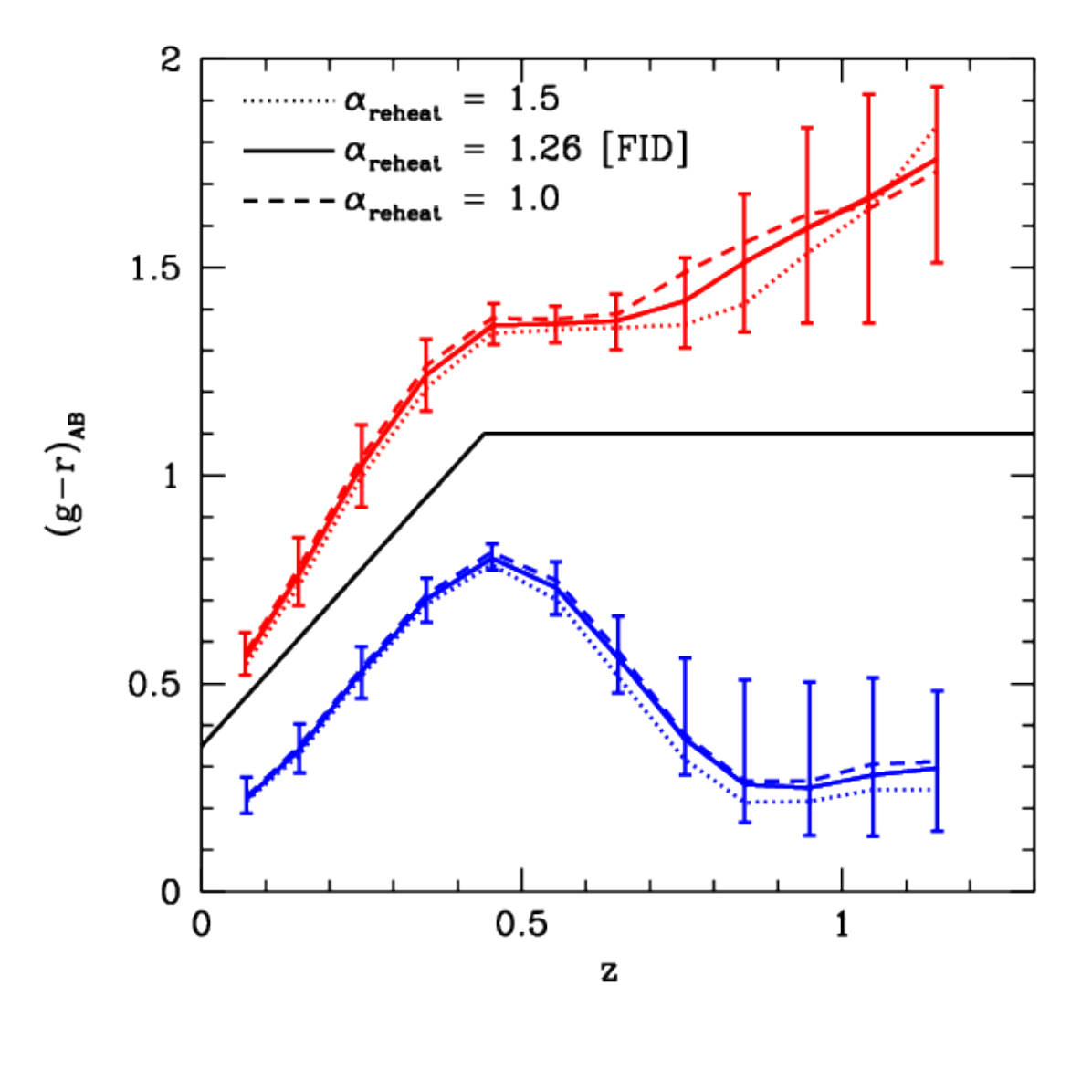} }%
    {\includegraphics[width=0.48\textwidth]{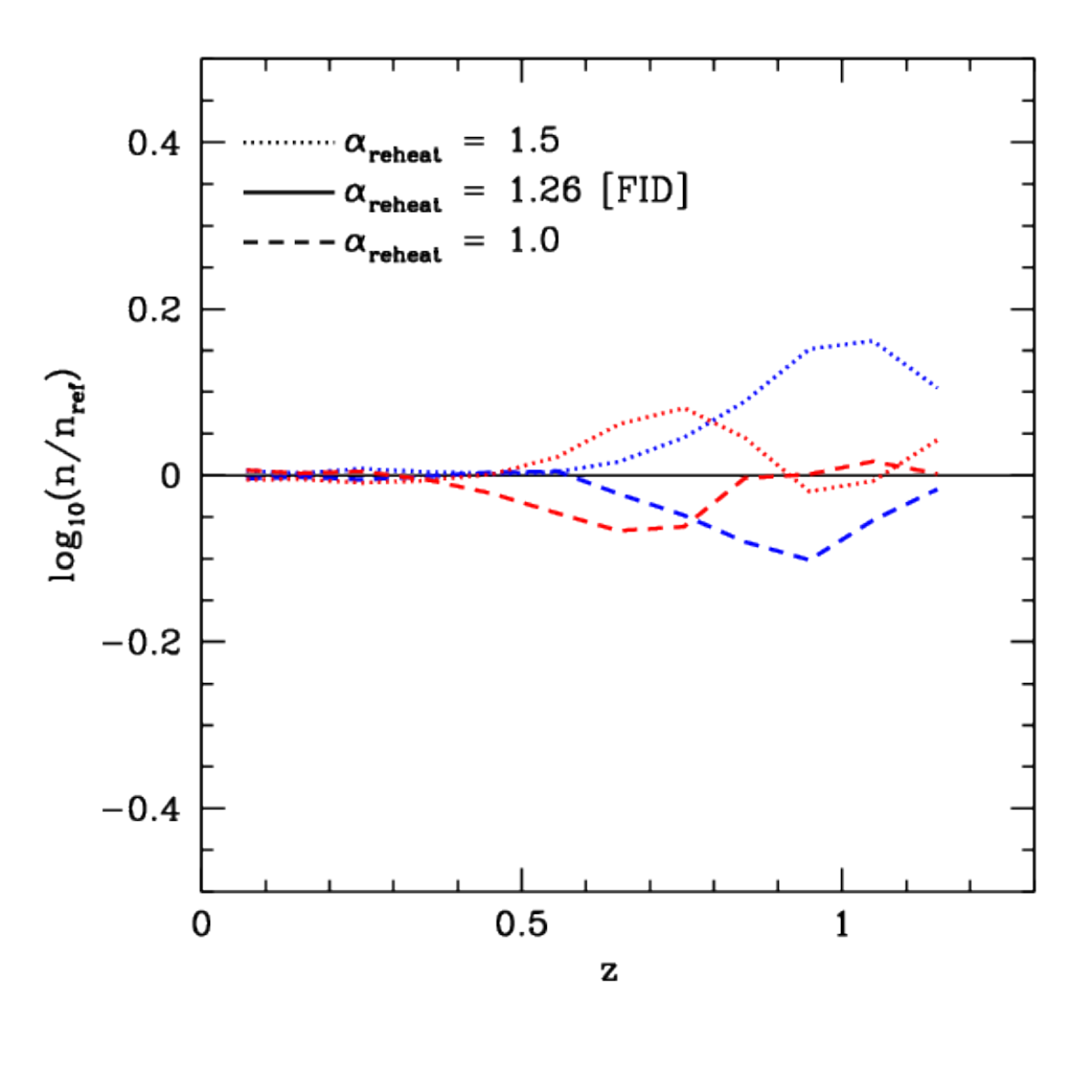} }%
\caption{The impact on galaxy colours of changing $\alpha_{\rm reheat}$, the parameter that controls the timescale for gas heated by SNe to be reincorporated into the hot halo, so that it can be considered for cooling. The value in the fiducial model is $\alpha_{\rm reheat}=1.26$: dotted lines show the results for $\alpha_{\rm reheat}=1.5$ and dashed lines show $\alpha_{\rm reheat}=1.0$. The top panel shows the median $g-r$ colour versus redshift, along with the 25-75 percentile range. The bottom panel shows the log of the ratio between the number of galaxies in the variant model and the number of galaxies in the fiducial model for the desired population (red or blue).}%
    \label{fig:gr_var1}%
\end{figure}
\begin{figure}
\centering
    {\includegraphics[width=0.48\textwidth]{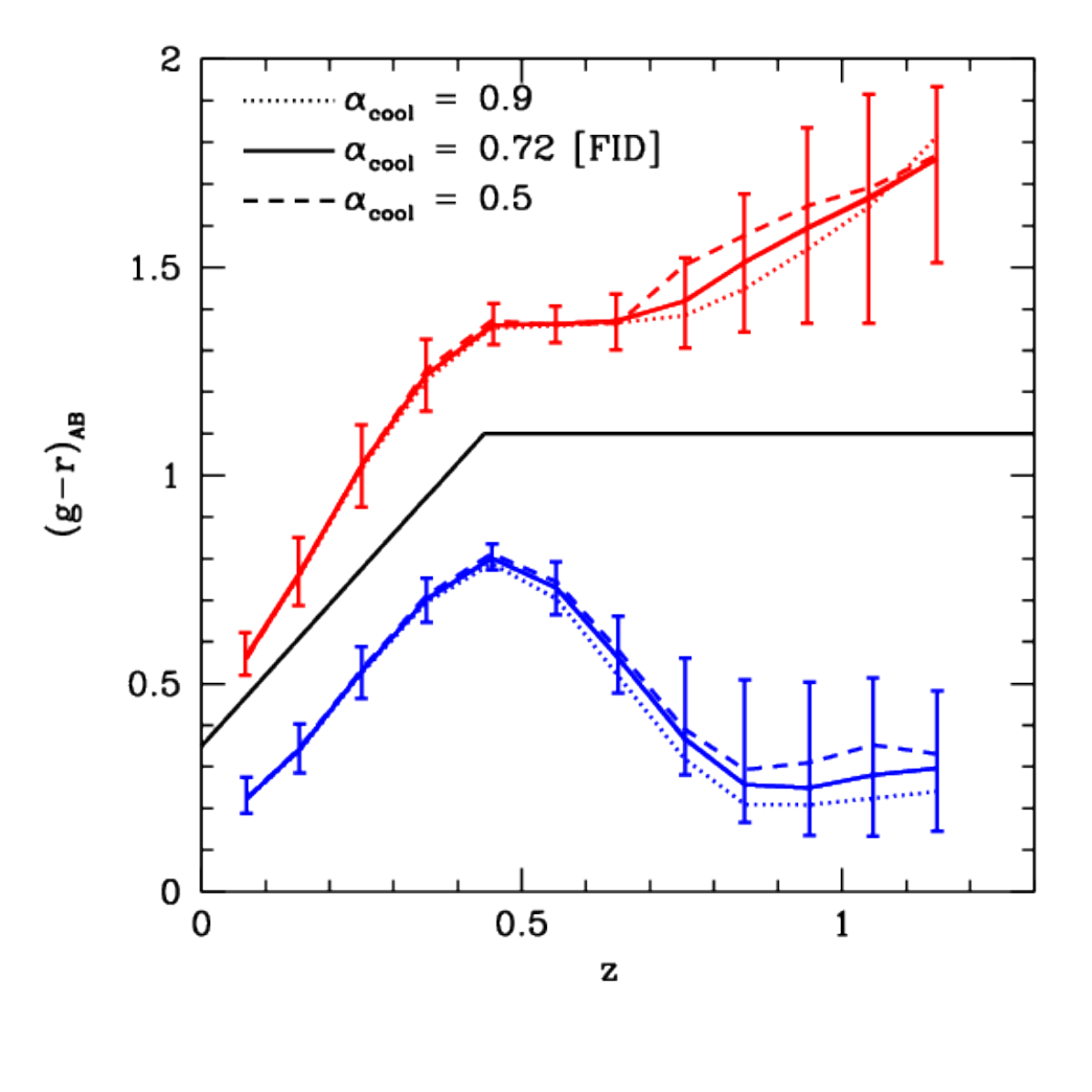} }%
    {\includegraphics[width=0.48\textwidth]{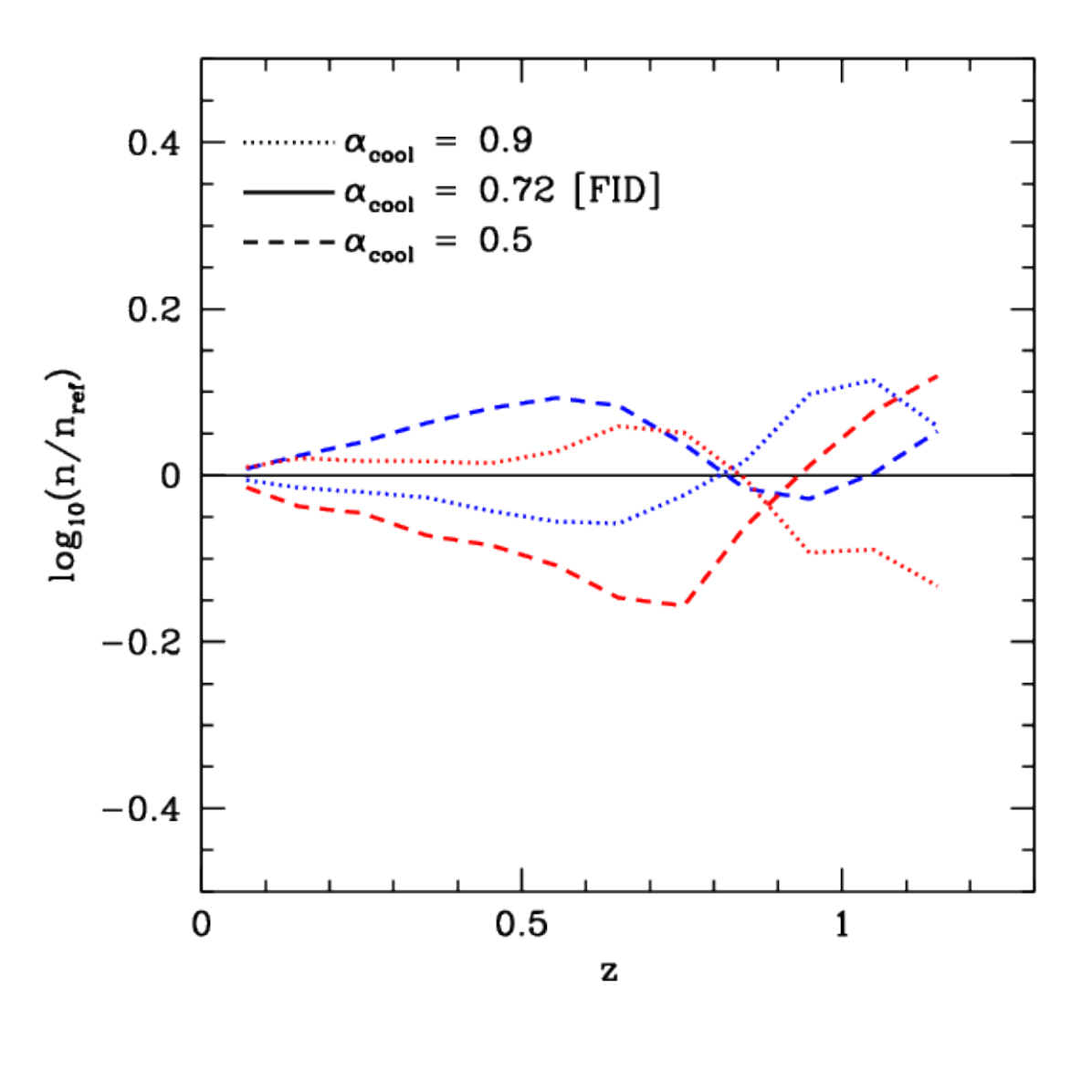} }%
\caption{The effect on galaxy colours of changing the parameter that governs the halo mass at which the AGN heating starts to prevent the cooling of the gas, $\alpha_{\rm cool}$. The top panel shows the median $g-r$ colour versus redshift, along with the 25-75 percentile range. The value in the fiducial model is $\alpha_{\rm cool}=0.72$. The dashed lines show the model with $\alpha_{\rm cool}=0.5$ and the dotted lines show the results for $\alpha_{\rm cool}=0.9$. With the same colour and line scheme as the upper panel, the bottom panel shows the log of the ratio between the number of galaxies in the variant model and the number of galaxies in the fiducial model for the population being studied (red or blue).}%
    \label{fig:gr_var3}%
\end{figure}
From Fig.~\ref{fig:gr_var}~to~\ref{fig:gr_var3} we show the model predictions for the median observer frame colours in the top panel and in the bottom panel we show the change in the number density of galaxies in the red and blue populations as a function of redshift. 
Specifically, in each figure, the upper panel shows the observer frame colours as a function of redshift, with the solid line showing the fiducial model and the dotted and dashed lines showing the predictions for the rescaled variants; dashed lines show the predictions for the lower value of the parameter varied and dotted lines the higher value. We leave for reference the black line indicating the separation used to classify red and blue galaxies in the colour - redshift plane (See Eqn~\ref{eq:redblue}).
The lower panel shows the logarithm of the ratio between the number of galaxies in the red or blue populations, labelled as $n$, and the number of objects for the same population in the fiducial model, $n_{\rm{ref}}$. We draw a horizontal line at $\log_{10}(n/n_{\rm{ref}})=0$ as this would be the place where the lines would lay in case the number of galaxies per population is not altered from the fiducial model.

The largest change in the median colours is found after changing the strength of the SNe feedback parameter, $V_{\rm SN}$, as shown in Fig.~\ref{fig:gr_var}. 
The $g-r$ colour shifts by more than the inter-quartile range of the model predictions on perturbing the SNe feedback. 
As well as the shift in the median colours, there are appreciable changes, of up to a factor of three, in the number of objects in the red and blue populations for this parameter change. 
From Fig.~\ref{fig:gr_var2}, instead, we learn that perturbing the star formation efficiency, $\nu_{\rm SF}$, results in only a small shift in the predicted median colours for red galaxies but a larger shift in the median colours of blue galaxies. 
The number of galaxies in the red and blue populations changes, by up to a factor of two. 
Fig.~\ref{fig:gr_var1} shows that the median colours of the red and blue populations hardly change on perturbing $\alpha_{\rm reheat}$. 
The change in the number of objects in these populations is modest. 
Finally, Fig.~\ref{fig:gr_var3} shows that the median colours of the red and blue populations are fairly insensitive to the value of $\alpha_{\rm cool}$ until $z \sim 1$. 
The changes in number density for this parameter change are also small.

In conclusion, we can state that in our models, the strength of the supernovae feedback, as controlled by the parameter $V_{\rm{SN}}$, is the physical process that alters the most the location of the red and blue population in the colour - redshift plots studied here (see upper panel of Fig.~\ref{fig:gr_var}). 
As a consequence, we can see that the population of red and blue galaxies change significantly in numbers (bottom panel of Fig.~\ref{fig:gr_var}) as the suppression of star formation is directly related to colours. Nevertheless, the overall trend of the colour - redshift relation is preserved making this test a good candidate for testing the accuracy of galaxy formation models.

\section{Conclusions}
\label{sec:conclusions}

We have presented a new observational test of galaxy formation models using a novel narrow band imaging survey, PAUS \citep{eriksen19,padilla19}. 
The narrow band imaging provides highly accurate photometric redshifts, which allow us to measure how galaxy properties evolve with redshift. The use of photometric redshifts removes any potential biases associated with the successful measurement of spectroscopic redshifts, and allows us to quantify the evolution of galaxy colours over an unprecedented baseline in redshift for a single survey with a homogeneous selection. 
We focus on observer frame galaxy colours to minimise the model dependent processing that needs to be applied to the data. 
Hence, we do not need to model the $k$-correction needed to estimate a rest-frame magnitude from the observed photometry. 

The PAUS sample used here is magnitude limited to $i_{\rm AB} = 22.5$, with galaxy redshifts that are mainly distributed between $0<z<1.2$ with a peak occurring at about $z\sim0.5$ (see Fig.~\ref{fig:n_of_z}). 
Over this redshift range a significant change in the global star formation rate per unit volume is observed \citep{madau_dickinson_14}.

We focus on the observed $g-r$ colour and its evolution with redshift. The observed colour distribution shows a clear division into red and blue populations (as shown in Fig.~\ref{fig:colour_redshift_lightcone}). 
The observed colours evolve strongly with redshift. 
This is driven mostly by the redshifting of the spectral energy distribution of the galaxies, which means that the filters sample different absorption features with increasing redshift. 
A secondary driver of the colour evolution is the change in the intrinsic galaxy properties with redshift, such as the overall increase in the global SFR with increasing redshift.

Hence to compare theoretical predictions to the observations, it is necessary to model the bandshifting effects on the galaxy spectra energy distribution and to build a mock catalogue on an observer's past lightcone, rather than focusing on fixed redshift outputs \citep{Baugh:2008}. 
We do this by implementing the \galf semi-analytical model of galaxy formation into the P-Millennium N-body simulation, using one of the recalibrated models presented in \cite{Baugh:2019}. 
The construction of a lightcone mock catalogue is described in \cite{merson13}. An earlier PAUS mock was made using this approach by \cite{stothert18}, but with a different N-body simulation. 
The mass resolution in the P-Millennium N-body simulation is almost an order of magnitude better than that in the simulation available to \cite{stothert18}, allowing intrinsically fainter galaxies to be included in the mock. 
This allows the mock to recover more of the expected galaxies, particularly at low redshift. Also, the P-Millennium has four times as many snapshots as the previous simulation, which means that the calculation of galaxy positions and magnitudes is more accurate than before. 
This is because having a higher number of redshift outputs in the same redshift range, hence more binned, facilitates the interpolation of properties between them.

The galaxy formation model used to build the mock is calibrated against mostly local observations. In particular, \cite{Baugh:2019} focused on the reproduction of the optical $b_{\rm J}$-band luminosity function and the HI mass function in the recalibration of the model parameters (the recalibration was necessary because of the change of cosmology in the P-Millennium, compared with earlier runs, and the improvement in the mass resolution). Hence, a useful entry level test of the model is that it reproduces the number counts in the PAUS survey as a function of apparent magnitude and redshift. 

The observed number counts are reproduced closely by the mock catalogue (Fig.~\ref{fig:num_counts}). This exercise also showed the importance of a robust and accurate algorithm for star-galaxy separation, in order to make a reliable comparison of galaxy counts with the model. 
This is particularly relevant at bright apparent magnitudes where stars make up a larger fraction of the total counts of objects. 
We also investigated if the number counts of galaxies change when we restrict our attention to galaxies with an estimated photometric redshift. 
The reason for this test is that to have an estimate for the photometric redshift, the requirement is to have that galaxy observed in at least 30 of the 40 narrow band filters, and not all galaxies in the PAUS W1 and W3 fields meet this criterion. 
Because of this, we want to make sure that using galaxies with an estimated photometric redshift is not introducing any bias. 
In the first instance, when a galaxy has a photometric redshift estimated, the shape of the number counts is unchanged. 
However, there is a small reduction in amplitude and this can be taken into account by introducing a constant sampling factor that accommodates for the fraction of missing objects.
If we restrict attention to galaxies which, based on the quality parameter (see \citealt{eriksen19}), are inferred to have good photometric redshifts, the shape of the number counts changes, with the fraction of galaxies with high quality photometric redshifts varying strongly with apparent magnitude. 
This is an important result that must be considered any time that we use the quality parameter to select galaxies with good photometric redshifts to perform any statistical analysis. 
We note that although the shape of the number counts is altered by retaining only those galaxies which are believed to have high quality photometric redshifts (comparing the brown line to the blue line in Fig.~\ref{fig:num_counts}), the colour distribution is not altered (as can be seen by comparing the top and bottom panels of Fig.~\ref{fig:gr_histo} which is selected over a narrow range in apparent magnitude close to  $i_{\rm{AB}}\sim 22$). This implies that the colour magnitude relation is flat at faint apparent magnitudes. 
Finally, another test that ensures us about the ability of the model to reproduce the observations is the good match to the overall galaxy redshift distribution, limited to the GAMA or PAUS survey apparent magnitude cuts. 

With the aim of testing further our model, we use the clear separation between galaxies in the colour - redshift plane (Fig.~\ref{fig:colour_redshift_lightcone}) to divide galaxies into red and blue populations. 
This definition works well for both the PAUS observations and the \galf mock catalogue. 
Reassuringly, when we limit our attention to those galaxies with high quality photometric redshifts in the observations, the colour distribution does not change, unlike the overall galaxy counts. 
The observer frame colour redshift relation from a photometric redshift survey like PAUS is therefore statistically robust to test galaxy formation models. 
Qualitatively, the colour-redshift plane looks similar in the model and observations. 
The red and blue populations are more sharply defined in the model than in the observations. 
This bimodality is greatly reduced if we include photometric errors in the model galaxy magnitudes, at the level expected for point sources, which implies that our model for the photometric errors may overestimate the errors.
There is good agreement between the median colours (and interquartile range) of the red and blue galaxies as a function of redshift. 
PAUS is able to probe the colour - redshift relation over a wide baseline in redshift (from $z=0$ to $z=1.2$) with a homogeneous selection.

We also look at the distribution of the observed colour $g-r$ for an apparent magnitude selected subset of the galaxies in redshift bins (Fig.~\ref{fig:gr_histo}).
Again, this test seems to be unaffected when only considering the $50$ per cent of galaxies with the best quality redshifts. 
The comparison between the model and the data is good at low redshifts. At high redshift, the bimodality in colours is stronger in the model than in the observations when we use the unperturbed magnitudes outputted by the model. 
However, this discrepancy is greatly reduced once photometric errors are included in the \galf predictions.

Finally, we examine the sensitivity of the model predictions to perturbations in the values of several key model parameters. 
These changes can alter the median colours of the red and blue populations and the number of galaxies in each population. 
For most of the parameter changes we considered, the median colours were unchanged, with small changes in the number of galaxies in the red and blue clouds. 
The parameter that controls the mass loading of supernovae-driven winds does produce a noticeable change in both the median colours and the number of galaxies in the red and blue populations, suggesting that the observed colours could be used as a further constraint on this model parameter. According to our model the fiducial value of $V_{\rm{SN}}= 380\, {\rm km/s}$ is the one reproducing better the observations in the colour - redshift plane (see a full discussion of this parameter in section~3.5.2 of \citealt{lacey:2016}).  

Although there is still some room for improvement in the accuracy of \galf predictions for galaxy colours, the tests presented here are mostly satisfied by the \galf model and they seem to be a good indicator of the accuracy of the model predictions for future galaxy surveys, over a key epoch in the history of galaxy evolution.

\section*{acknowledgements}
GM was supported by a PhD studentship with the Durham Centre for Doctoral Training in Data Intensive Science, funded by the UK Science and Technology Facilities Council (STFC, ST/P006744/1) and Durham University.
GM is now supported by the Collaborative Research Fund under Grant No. C6017-20G which is issued by the Research Grants Council of Hong Kong S.A.R.
CMB and PN acknowledge support from the STFC through ST/T000244/1.
PF acknowledges support from Ministerio de Ciencia e Innovacion, project PID2019-111317GB-C31, the European Research Executive Agency HORIZON-MSCA-2021-SE-01 Research and Innovation programme under the Marie Skłodowska-Curie grant agreement number 101086388 (LACEGAL). PF is also partially supported by the program Unidad de Excelencia María de Maeztu CEX2020-001058-M.
This work used the DiRAC@Durham facility managed by the Institute for Computational Cosmology on behalf of the STFC DiRAC HPC Facility (www.dirac.ac.uk). The equipment was funded by BEIS capital funding via STFC capital grants ST/K00042X/1, ST/P002293/1, ST/R002371/1 and ST/S002502/1, Durham University and STFC operations grant ST/R000832/1. DiRAC is part of the National e-Infrastructure. 
The PAU Survey is partially supported by MINECO under grants CSD2007-00060, AYA2015-71825, ESP2017-89838, PGC2018-094773, PGC2018-102021, PID2019-111317GB, SEV-2016-0588, SEV-2016-0597, MDM-2015-0509 and Juan de la Cierva fellowship and LACEGAL and EWC Marie Sklodowska-Curie grant No 734374 and no.776247 from the EU Horizon 2020 Programme. IEEC and IFAE are partially funded by the CERCA and Beatriu de Pinos program of the Generalitat de Catalunya. Funding for PAUS has also been provided by Durham University (via the ERC StG DEGAS-259586), ETH Zurich, Leiden University (via ERC StG ADULT-279396 and Netherlands Organisation for Scientific Research (NWO) Vici grant 639.043.512), University College London and from the European Union's Horizon 2020 research and innovation programme under the grant agreement No 776247 EWC. The PAU data center is hosted by the Port d'Informaci\'o Cient\'ifica (PIC), maintained through a collaboration of CIEMAT and IFAE, with additional support from Universitat Aut\`onoma de Barcelona and European Regional Development Fund. We acknowledge the PIC services department team for their support and fruitful discussions.
JYHS acknowledges financial support via the Fundamental Research Grant Scheme (FRGS) by the Malaysian Ministry of Higher Education with code FRGS/1/2023/STG07/USM/02/14.
PR acknowledges the support by the Tsinghua Shui Mu Scholarship, the funding of the National Key R\&D Program of China (grant no. 2018YFA0404503), the National Science Foundation of China (grant no. 12073014), the science research grants from the China Manned Space Project with No. CMS-CSST2021-A05, and the Tsinghua University Initiative Scientific Research Program (No. 20223080023).
JGB acknowledges support from the Spanish Research Project PID2021-123012NB-C43 [MICINNFEDER], and the Centro de Excelencia Severo Ochoa Program CEX2020-001007-S at IFT.
HHo acknowledges support from the Netherlands Organisation for Scientific Research (NWO) through grant 639.043.512.
HHi is supported by a DFG Heisenberg grant (Hi 1495/5-1), the DFG Collaborative Research Center SFB1491, as well as an ERC Consolidator Grant (No. 770935).
JC acknowledges support from the Spanish Research Project PID2021-123012NA-C44 [MICINNFEDER].
MS has been supported by the Polish National Agency for Academic Exchange (Bekker grant BPN/BEK/2021/1/00298/DEC/1), the European Union's Horizon 2020 Research and Innovation programme under the Maria Sklodowska-Curie grant agreement (No. 754510).
EJG was partially supported by Ministerio de Ciencia e Innovación, Agencia Estatal de Investigación and FEDER (PID2021-123012NA-C44). I finally want to thank all the informal contributions from the GALFORM team spread around the world including Shaun Cole, Cedric Lacey, Piotr Oleskiewicz, Andrew Griffin, Alex Smith, Adarsh Kumar, Difu Shi, Sownak Bose, Andrew Benson, Will Cowley, Ed Elliot, Alex Merson, Violeta Gonzalez-Perez, the whole COSMA SUPPORT team including Alan Lotts, Lydia Heck, and the whole PAUS collaboration including Santi Serrano, Alex Alarcon and Andrea Pocino.
%

\section*{Data Availability}
The data that support the findings of this study are stored at the Durham COSMA facilities (GALFORM mock) and at the Barcelona PIC facilities (PAUS observations). The GALFORM mock is available from the corresponding author, GM, or the Durham GALFORM team, while for the observation please contact the PAUS team through the website \url{https://pausurvey.org/}. In particular, the versions of the data used in this work are PAUS photometric production 972 for the W3 field and 979 for the W1 field.

\bibliographystyle{mnras}
\bibliography{manzoni}

\begin{thebibliography}{}
\makeatletter
\relax
\def\mn@urlcharsother{\let\do\@makeother \do\$\do\&\do\#\do\^\do\_\do\%\do\~}
\def\mn@doi{\begingroup\mn@urlcharsother \@ifnextchar [ {\mn@doi@}
  {\mn@doi@[]}}
\def\mn@doi@[#1]#2{\def\@tempa{#1}\ifx\@tempa\@empty \href
  {http://dx.doi.org/#2} {doi:#2}\else \href {http://dx.doi.org/#2} {#1}\fi
  \endgroup}
\def\mn@eprint#1#2{\mn@eprint@#1:#2::\@nil}
\def\mn@eprint@arXiv#1{\href {http://arxiv.org/abs/#1} {{\tt arXiv:#1}}}
\def\mn@eprint@dblp#1{\href {http://dblp.uni-trier.de/rec/bibtex/#1.xml}
  {dblp:#1}}
\def\mn@eprint@#1:#2:#3:#4\@nil{\def\@tempa {#1}\def\@tempb {#2}\def\@tempc
  {#3}\ifx \@tempc \@empty \let \@tempc \@tempb \let \@tempb \@tempa \fi \ifx
  \@tempb \@empty \def\@tempb {arXiv}\fi \@ifundefined
  {mn@eprint@\@tempb}{\@tempb:\@tempc}{\expandafter \expandafter \csname
  mn@eprint@\@tempb\endcsname \expandafter{\@tempc}}}

\bibitem[\protect\citeauthoryear{{Alarcon} et~al.,}{{Alarcon}
  et~al.}{2021}]{alarcon21}
{Alarcon} A.,  et~al., 2021, \mn@doi [\mnras] {10.1093/mnras/staa3659}, \href
  {https://ui.adsabs.harvard.edu/abs/2021MNRAS.501.6103A} {501, 6103}

\bibitem[\protect\citeauthoryear{{Baugh}}{{Baugh}}{2006}]{baugh06}
{Baugh} C.~M.,  2006, \mn@doi [Reports on Progress in Physics]
  {10.1088/0034-4885/69/12/R02}, \href
  {http://adsabs.harvard.edu/abs/2006RPPh...69.3101B} {69, 3101}

\bibitem[\protect\citeauthoryear{{Baugh}}{{Baugh}}{2008}]{Baugh:2008}
{Baugh} C.~M.,  2008, \mn@doi [Philosophical Transactions of the Royal Society
  of London Series A] {10.1098/rsta.2008.0192}, \href
  {https://ui.adsabs.harvard.edu/abs/2008RSPTA.366.4381B} {366, 4381}

\bibitem[\protect\citeauthoryear{{Baugh} \& {Efstathiou}}{{Baugh} \&
  {Efstathiou}}{1993}]{Baugh:1993}
{Baugh} C.~M.,  {Efstathiou} G.,  1993, \mn@doi [\mnras]
  {10.1093/mnras/265.1.145}, \href
  {https://ui.adsabs.harvard.edu/abs/1993MNRAS.265..145B} {265, 145}

\bibitem[\protect\citeauthoryear{{Baugh}, {Lacey}, {Frenk}, {Granato}, {Silva},
  {Bressan}, {Benson}  \& {Cole}}{{Baugh} et~al.}{2005}]{Baugh:2005}
{Baugh} C.~M.,  {Lacey} C.~G.,  {Frenk} C.~S.,  {Granato} G.~L.,  {Silva} L.,
  {Bressan} A.,  {Benson} A.~J.,   {Cole} S.,  2005, \mn@doi [\mnras]
  {10.1111/j.1365-2966.2004.08553.x}, \href
  {https://ui.adsabs.harvard.edu/abs/2005MNRAS.356.1191B} {356, 1191}

\bibitem[\protect\citeauthoryear{{Baugh} et~al.,}{{Baugh}
  et~al.}{2019}]{Baugh:2019}
{Baugh} C.~M.,  et~al., 2019, \mn@doi [\mnras] {10.1093/mnras/sty3427}, \href
  {https://ui.adsabs.harvard.edu/abs/2019MNRAS.483.4922B} {483, 4922}

\bibitem[\protect\citeauthoryear{{Baugh}, {Lacey}, {Gonzalez-Perez}  \&
  {Manzoni}}{{Baugh} et~al.}{2022}]{Baugh:2022}
{Baugh} C.~M.,  {Lacey} C.~G.,  {Gonzalez-Perez} V.,   {Manzoni} G.,  2022,
  \mn@doi [\mnras] {10.1093/mnras/stab3506}, \href
  {https://ui.adsabs.harvard.edu/abs/2022MNRAS.510.1880B} {510, 1880}

\bibitem[\protect\citeauthoryear{{Benson}}{{Benson}}{2010}]{Benson:2010}
{Benson} A.~J.,  2010, \mn@doi [\physrep] {10.1016/j.physrep.2010.06.001},
  \href {https://ui.adsabs.harvard.edu/abs/2010PhR...495...33B} {495, 33}

\bibitem[\protect\citeauthoryear{{Benson}, {Baugh}, {Cole}, {Frenk}  \&
  {Lacey}}{{Benson} et~al.}{2000}]{Benson:00}
{Benson} A.~J.,  {Baugh} C.~M.,  {Cole} S.,  {Frenk} C.~S.,   {Lacey} C.~G.,
  2000, \mn@doi [\mnras] {10.1046/j.1365-8711.2000.03470.x}, \href
  {https://ui.adsabs.harvard.edu/abs/2000MNRAS.316..107B} {316, 107}

\bibitem[\protect\citeauthoryear{{Blaizot}, {Wadadekar}, {Guiderdoni},
  {Colombi}, {Bertin}, {Bouchet}, {Devriendt}  \& {Hatton}}{{Blaizot}
  et~al.}{2005}]{Blaizot:2005}
{Blaizot} J.,  {Wadadekar} Y.,  {Guiderdoni} B.,  {Colombi} S.~T.,  {Bertin}
  E.,  {Bouchet} F.~R.,  {Devriendt} J. E.~G.,   {Hatton} S.,  2005, \mn@doi
  [\mnras] {10.1111/j.1365-2966.2005.09019.x}, \href
  {https://ui.adsabs.harvard.edu/abs/2005MNRAS.360..159B} {360, 159}

\bibitem[\protect\citeauthoryear{{Blitz} \& {Rosolowsky}}{{Blitz} \&
  {Rosolowsky}}{2006}]{Blitz:2006}
{Blitz} L.,  {Rosolowsky} E.,  2006, \mn@doi [\apj] {10.1086/505417}, \href
  {https://ui.adsabs.harvard.edu/abs/2006ApJ...650..933B} {650, 933}

\bibitem[\protect\citeauthoryear{{Bower}, {Benson}, {Malbon}, {Helly}, {Frenk},
  {Baugh}, {Cole}  \& {Lacey}}{{Bower} et~al.}{2006}]{bower:2006}
{Bower} R.~G.,  {Benson} A.~J.,  {Malbon} R.,  {Helly} J.~C.,  {Frenk} C.~S.,
  {Baugh} C.~M.,  {Cole} S.,   {Lacey} C.~G.,  2006, \mn@doi [\mnras]
  {10.1111/j.1365-2966.2006.10519.x}, \href
  {http://adsabs.harvard.edu/abs/2006MNRAS.370..645B} {370, 645}

\bibitem[\protect\citeauthoryear{{Boylan-Kolchin}, {Springel}, {White},
  {Jenkins}  \& {Lemson}}{{Boylan-Kolchin} et~al.}{2009}]{Boylan-Kolchin:2009}
{Boylan-Kolchin} M.,  {Springel} V.,  {White} S. D.~M.,  {Jenkins} A.,
  {Lemson} G.,  2009, \mn@doi [\mnras] {10.1111/j.1365-2966.2009.15191.x},
  \href {https://ui.adsabs.harvard.edu/abs/2009MNRAS.398.1150B} {398, 1150}

\bibitem[\protect\citeauthoryear{{Brammer}, {van Dokkum}  \& {Coppi}}{{Brammer}
  et~al.}{2008}]{Brammer:2008}
{Brammer} G.~B.,  {van Dokkum} P.~G.,   {Coppi} P.,  2008, \mn@doi [\apj]
  {10.1086/591786}, \href
  {https://ui.adsabs.harvard.edu/abs/2008ApJ...686.1503B} {686, 1503}

\bibitem[\protect\citeauthoryear{{Bravo}, {Lagos}, {Robotham}, {Bellstedt}  \&
  {Obreschkow}}{{Bravo} et~al.}{2020}]{Bravo2020}
{Bravo} M.,  {Lagos} C. d.~P.,  {Robotham} A. S.~G.,  {Bellstedt} S.,
  {Obreschkow} D.,  2020, \mn@doi [\mnras] {10.1093/mnras/staa2027}, \href
  {https://ui.adsabs.harvard.edu/abs/2020MNRAS.497.3026B} {497, 3026}

\bibitem[\protect\citeauthoryear{{Cabayol} et~al.,}{{Cabayol}
  et~al.}{2021}]{cabayol21}
{Cabayol} L.,  et~al., 2021, \mn@doi [\mnras] {10.1093/mnras/stab1909}, \href
  {https://ui.adsabs.harvard.edu/abs/2021MNRAS.506.4048C} {506, 4048}

\bibitem[\protect\citeauthoryear{{Cabayol} et~al.,}{{Cabayol}
  et~al.}{2023}]{Cabayol:2023}
{Cabayol} L.,  et~al., 2023, \mn@doi [\aap] {10.1051/0004-6361/202245027},
  \href {https://ui.adsabs.harvard.edu/abs/2023A&A...671A.153C} {671, A153}

\bibitem[\protect\citeauthoryear{{Capak} et~al.,}{{Capak}
  et~al.}{2007}]{capak07}
{Capak} P.,  et~al., 2007, \mn@doi [\apjs] {10.1086/519081}, \href
  {https://ui.adsabs.harvard.edu/abs/2007ApJS..172...99C} {172, 99}

\bibitem[\protect\citeauthoryear{{Casas} et~al.,}{{Casas}
  et~al.}{2016}]{casas16}
{Casas} R.,  et~al., 2016, in {Evans} C.~J.,  {Simard} L.,   {Takami} H.,  eds,
   Society of Photo-Optical Instrumentation Engineers (SPIE) Conference Series
  Vol. 9908, Ground-based and Airborne Instrumentation for Astronomy VI. p.
  99084K, \mn@doi{10.1117/12.2232422}

\bibitem[\protect\citeauthoryear{{Cole}, {Lacey}, {Baugh}  \& {Frenk}}{{Cole}
  et~al.}{2000}]{cole:2000}
{Cole} S.,  {Lacey} C.~G.,  {Baugh} C.~M.,   {Frenk} C.~S.,  2000, \mn@doi
  [\mnras] {10.1046/j.1365-8711.2000.03879.x}, \href
  {http://adsabs.harvard.edu/abs/2000MNRAS.319..168C} {319, 168}

\bibitem[\protect\citeauthoryear{{Conroy}}{{Conroy}}{2013}]{Conroy2013}
{Conroy} C.,  2013, \mn@doi [\araa] {10.1146/annurev-astro-082812-141017},
  \href {https://ui.adsabs.harvard.edu/abs/2013ARA&A..51..393C} {51, 393}

\bibitem[\protect\citeauthoryear{{Cuillandre} et~al.,}{{Cuillandre}
  et~al.}{2012}]{CFHTLS:2012}
{Cuillandre} J.-C.~J.,  et~al., 2012, in {Peck} A.~B.,  {Seaman} R.~L.,
  {Comeron} F.,  eds,  Society of Photo-Optical Instrumentation Engineers
  (SPIE) Conference Series Vol. 8448, Observatory Operations: Strategies,
  Processes, and Systems IV. p. 84480M, \mn@doi{10.1117/12.925584}

\bibitem[\protect\citeauthoryear{{DESI Collaboration} et~al.,}{{DESI
  Collaboration} et~al.}{2016}]{DESI:2016}
{DESI Collaboration} et~al., 2016, arXiv e-prints, \href
  {https://ui.adsabs.harvard.edu/abs/2016arXiv161100036D} {p. arXiv:1611.00036}

\bibitem[\protect\citeauthoryear{{DESI Collaboration} et~al.,}{{DESI
  Collaboration} et~al.}{2022}]{DESI:2022}
{DESI Collaboration} et~al., 2022, \mn@doi [\aj] {10.3847/1538-3881/ac882b},
  \href {https://ui.adsabs.harvard.edu/abs/2022AJ....164..207D} {164, 207}

\bibitem[\protect\citeauthoryear{{Daddi} et~al.,}{{Daddi}
  et~al.}{2007}]{Daddi:2007}
{Daddi} E.,  et~al., 2007, \mn@doi [\apj] {10.1086/521818}, \href
  {https://ui.adsabs.harvard.edu/abs/2007ApJ...670..156D} {670, 156}

\bibitem[\protect\citeauthoryear{{Davies} et~al.,}{{Davies}
  et~al.}{2018}]{Davies:2018}
{Davies} L.~J.~M.,  et~al., 2018, \mn@doi [\mnras] {10.1093/mnras/sty1553},
  \href {https://ui.adsabs.harvard.edu/abs/2018MNRAS.480..768D} {480, 768}

\bibitem[\protect\citeauthoryear{{Driver} et~al.,}{{Driver}
  et~al.}{2009}]{driver09}
{Driver} S.~P.,  et~al., 2009, \mn@doi [Astronomy and Geophysics]
  {10.1111/j.1468-4004.2009.50512.x}, \href
  {https://ui.adsabs.harvard.edu/abs/2009A&G....50e..12D} {50, 5.12}

\bibitem[\protect\citeauthoryear{{Driver} et~al.,}{{Driver}
  et~al.}{2011}]{Driver:2011}
{Driver} S.~P.,  et~al., 2011, \mn@doi [\mnras]
  {10.1111/j.1365-2966.2010.18188.x}, \href
  {https://ui.adsabs.harvard.edu/abs/2011MNRAS.413..971D} {413, 971}

\bibitem[\protect\citeauthoryear{{Elliott}, {Baugh}  \& {Lacey}}{{Elliott}
  et~al.}{2021}]{elliott21}
{Elliott} E.~J.,  {Baugh} C.~M.,   {Lacey} C.~G.,  2021, \mn@doi [\mnras]
  {10.1093/mnras/stab1837}, \href
  {https://ui.adsabs.harvard.edu/abs/2021MNRAS.506.4011E} {506, 4011}

\bibitem[\protect\citeauthoryear{{Erben} et~al.,}{{Erben}
  et~al.}{2013}]{Erben:2013}
{Erben} T.,  et~al., 2013, \mn@doi [\mnras] {10.1093/mnras/stt928}, \href
  {https://ui.adsabs.harvard.edu/abs/2013MNRAS.433.2545E} {433, 2545}

\bibitem[\protect\citeauthoryear{{Eriksen} et~al.,}{{Eriksen}
  et~al.}{2019}]{eriksen19}
{Eriksen} M.,  et~al., 2019, \mn@doi [\mnras] {10.1093/mnras/stz204}, \href
  {https://ui.adsabs.harvard.edu/abs/2019MNRAS.484.4200E} {484, 4200}

\bibitem[\protect\citeauthoryear{{Eriksen} et~al.,}{{Eriksen}
  et~al.}{2020}]{Eriksen2020}
{Eriksen} M.,  et~al., 2020, \mn@doi [\mnras] {10.1093/mnras/staa2265}, \href
  {https://ui.adsabs.harvard.edu/abs/2020MNRAS.497.4565E} {497, 4565}

\bibitem[\protect\citeauthoryear{Gonzalez-Perez, Lacey, Baugh, Lagos, Helly,
  Campbell  \& Mitchell}{Gonzalez-Perez et~al.}{2014}]{gonzalez14}
Gonzalez-Perez V.,  Lacey C.~G.,  Baugh C.~M.,  Lagos C. D.~P.,  Helly J.,
  Campbell D. J.~R.,   Mitchell P.~D.,  2014, \mn@doi [Monthly Notices of the
  Royal Astronomical Society] {10.1093/mnras/stt2410}, 439, 264

\bibitem[\protect\citeauthoryear{{Gonz{\'a}lez}, {Lacey}, {Baugh}, {Frenk}  \&
  {Benson}}{{Gonz{\'a}lez} et~al.}{2009}]{gonzalez09}
{Gonz{\'a}lez} J.~E.,  {Lacey} C.~G.,  {Baugh} C.~M.,  {Frenk} C.~S.,
  {Benson} A.~J.,  2009, \mn@doi [\mnras] {10.1111/j.1365-2966.2009.15057.x},
  \href {http://adsabs.harvard.edu/abs/2009MNRAS.397.1254G} {397, 1254}

\bibitem[\protect\citeauthoryear{{Guo}, {White}, {Angulo}, {Henriques},
  {Lemson}, {Boylan-Kolchin}, {Thomas}  \& {Short}}{{Guo}
  et~al.}{2013}]{Guo:2013}
{Guo} Q.,  {White} S.,  {Angulo} R.~E.,  {Henriques} B.,  {Lemson} G.,
  {Boylan-Kolchin} M.,  {Thomas} P.,   {Short} C.,  2013, \mn@doi [\mnras]
  {10.1093/mnras/sts115}, \href
  {https://ui.adsabs.harvard.edu/abs/2013MNRAS.428.1351G} {428, 1351}

\bibitem[\protect\citeauthoryear{{Guzzo} et~al.,}{{Guzzo}
  et~al.}{2014}]{guzzo14}
{Guzzo} L.,  et~al., 2014, \mn@doi [\aap] {10.1051/0004-6361/201321489}, \href
  {http://adsabs.harvard.edu/abs/2014A%26A...566A.108G} {566, A108}

\bibitem[\protect\citeauthoryear{{Hahn} et~al.,}{{Hahn}
  et~al.}{2023}]{DESIBGS:2023}
{Hahn} C.,  et~al., 2023, \mn@doi [\aj] {10.3847/1538-3881/accff8}, \href
  {https://ui.adsabs.harvard.edu/abs/2023AJ....165..253H} {165, 253}

\bibitem[\protect\citeauthoryear{{Hogg}, {Baldry}, {Blanton}  \&
  {Eisenstein}}{{Hogg} et~al.}{2002}]{Hogg:2002}
{Hogg} D.~W.,  {Baldry} I.~K.,  {Blanton} M.~R.,   {Eisenstein} D.~J.,  2002,
  arXiv e-prints, \href {https://ui.adsabs.harvard.edu/abs/2002astro.ph.10394H}
  {pp astro--ph/0210394}

\bibitem[\protect\citeauthoryear{{Jiang}, {Helly}, {Cole}  \& {Frenk}}{{Jiang}
  et~al.}{2014}]{Jiang:2014}
{Jiang} L.,  {Helly} J.~C.,  {Cole} S.,   {Frenk} C.~S.,  2014, \mn@doi
  [\mnras] {10.1093/mnras/stu390}, \href
  {https://ui.adsabs.harvard.edu/abs/2014MNRAS.440.2115J} {440, 2115}

\bibitem[\protect\citeauthoryear{{Kasparova} et~al.,}{{Kasparova}
  et~al.}{2021}]{Kasparova:2021}
{Kasparova} A.,  et~al., 2021, arXiv e-prints, \href
  {https://ui.adsabs.harvard.edu/abs/2021arXiv211204864K} {p. arXiv:2112.04864}

\bibitem[\protect\citeauthoryear{{Kauffman}}{{Kauffman}}{1999}]{Kauffman:99}
{Kauffman} G.,  1999, in American Astronomical Society Meeting Abstracts. p.
  67.01

\bibitem[\protect\citeauthoryear{{Kitzbichler} \& {White}}{{Kitzbichler} \&
  {White}}{2007a}]{Kitzbichler:2007}
{Kitzbichler} M.~G.,  {White} S.~D.~M.,  2007a, \mn@doi [\mnras]
  {10.1111/j.1365-2966.2007.11458.x}, \href
  {https://ui.adsabs.harvard.edu/abs/2007MNRAS.376....2K} {376, 2}

\bibitem[\protect\citeauthoryear{{Kitzbichler} \& {White}}{{Kitzbichler} \&
  {White}}{2007b}]{KW2007}
{Kitzbichler} M.~G.,  {White} S.~D.~M.,  2007b, \mn@doi [\mnras]
  {10.1111/j.1365-2966.2007.11458.x}, \href
  {https://ui.adsabs.harvard.edu/abs/2007MNRAS.376....2K} {376, 2}

\bibitem[\protect\citeauthoryear{{Kuijken} et~al.,}{{Kuijken}
  et~al.}{2019}]{Kuijken:2019}
{Kuijken} K.,  et~al., 2019, \mn@doi [\aap] {10.1051/0004-6361/201834918},
  \href {https://ui.adsabs.harvard.edu/abs/2019A&A...625A...2K} {625, A2}

\bibitem[\protect\citeauthoryear{{Lacey} et~al.,}{{Lacey}
  et~al.}{2016}]{lacey:2016}
{Lacey} C.~G.,  et~al., 2016, \mn@doi [\mnras] {10.1093/mnras/stw1888}, \href
  {http://adsabs.harvard.edu/abs/2016MNRAS.462.3854L} {462, 3854}

\bibitem[\protect\citeauthoryear{{Lagos}, {Tobar}, {Robotham}, {Obreschkow},
  {Mitchell}, {Power}  \& {Elahi}}{{Lagos} et~al.}{2018}]{Lagos2018}
{Lagos} C. d.~P.,  {Tobar} R.~J.,  {Robotham} A. S.~G.,  {Obreschkow} D.,
  {Mitchell} P.~D.,  {Power} C.,   {Elahi} P.~J.,  2018, \mn@doi [\mnras]
  {10.1093/mnras/sty2440}, \href
  {https://ui.adsabs.harvard.edu/abs/2018MNRAS.481.3573L} {481, 3573}

\bibitem[\protect\citeauthoryear{{Lagos} et~al.,}{{Lagos}
  et~al.}{2019}]{Lagos:2019}
{Lagos} C. D.~P.,  et~al., 2019, \mn@doi [\mnras] {10.1093/mnras/stz2427},
  \href {https://ui.adsabs.harvard.edu/abs/2019MNRAS.tmp.2096L} {p.~2096}

\bibitem[\protect\citeauthoryear{{Laureijs} et~al.,}{{Laureijs}
  et~al.}{2011}]{Euclid:2011}
{Laureijs} R.,  et~al., 2011, arXiv e-prints, \href
  {https://ui.adsabs.harvard.edu/abs/2011arXiv1110.3193L} {p. arXiv:1110.3193}

\bibitem[\protect\citeauthoryear{{Loveday} et~al.,}{{Loveday}
  et~al.}{2015}]{Loveday:2015}
{Loveday} J.,  et~al., 2015, \mn@doi [\mnras] {10.1093/mnras/stv1013}, \href
  {https://ui.adsabs.harvard.edu/abs/2015MNRAS.451.1540L} {451, 1540}

\bibitem[\protect\citeauthoryear{{Madau} \& {Dickinson}}{{Madau} \&
  {Dickinson}}{2014}]{madau_dickinson_14}
{Madau} P.,  {Dickinson} M.,  2014, \mn@doi [\araa]
  {10.1146/annurev-astro-081811-125615}, \href
  {http://adsabs.harvard.edu/abs/2014ARA%26A..52..415M} {52, 415}

\bibitem[\protect\citeauthoryear{{Manzoni} et~al.,}{{Manzoni}
  et~al.}{2021}]{manzoni21}
{Manzoni} G.,  et~al., 2021, \mn@doi [\na] {10.1016/j.newast.2020.101515},
  \href {https://ui.adsabs.harvard.edu/abs/2021NewA...8401515M} {84, 101515}

\bibitem[\protect\citeauthoryear{{Merson} et~al.,}{{Merson}
  et~al.}{2013}]{merson13}
{Merson} A.~I.,  et~al., 2013, \mn@doi [\mnras] {10.1093/mnras/sts355}, \href
  {https://ui.adsabs.harvard.edu/abs/2013MNRAS.429..556M} {429, 556}

\bibitem[\protect\citeauthoryear{Navarro-Gironés et~al.,}{Navarro-Gironés
  et~al.}{2023}]{david23}
Navarro-Gironés D.,  et~al., 2023, The PAU Survey: Photometric redshift
  estimation in deep wide fields (\mn@eprint {arXiv} {2312.07581})

\bibitem[\protect\citeauthoryear{{Padilla} et~al.,}{{Padilla}
  et~al.}{2019}]{padilla19}
{Padilla} C.,  et~al., 2019, \mn@doi [\aj] {10.3847/1538-3881/ab0412}, \href
  {https://ui.adsabs.harvard.edu/abs/2019AJ....157..246P} {157, 246}

\bibitem[\protect\citeauthoryear{{Renard} et~al.,}{{Renard}
  et~al.}{2022}]{Renard:2022}
{Renard} P.,  et~al., 2022, \mn@doi [\mnras] {10.1093/mnras/stac1730}, \href
  {https://ui.adsabs.harvard.edu/abs/2022MNRAS.515..146R} {515, 146}

\bibitem[\protect\citeauthoryear{{Robotham}, {Bellstedt}, {Lagos}, {Thorne},
  {Davies}, {Driver}  \& {Bravo}}{{Robotham} et~al.}{2020}]{Robotham:2020}
{Robotham} A.~S.~G.,  {Bellstedt} S.,  {Lagos} C. d.~P.,  {Thorne} J.~E.,
  {Davies} L.~J.,  {Driver} S.~P.,   {Bravo} M.,  2020, \mn@doi [\mnras]
  {10.1093/mnras/staa1116}, \href
  {https://ui.adsabs.harvard.edu/abs/2020MNRAS.495..905R} {495, 905}

\bibitem[\protect\citeauthoryear{{Ruiz-Macias} et~al.,}{{Ruiz-Macias}
  et~al.}{2020}]{omar20}
{Ruiz-Macias} O.,  et~al., 2020, \mn@doi [Research Notes of the American
  Astronomical Society] {10.3847/2515-5172/abc25a}, \href
  {https://ui.adsabs.harvard.edu/abs/2020RNAAS...4..187R} {4, 187}

\bibitem[\protect\citeauthoryear{{Ruiz-Macias} et~al.,}{{Ruiz-Macias}
  et~al.}{2021}]{Omar:2021}
{Ruiz-Macias} O.,  et~al., 2021, \mn@doi [\mnras] {10.1093/mnras/stab292},
  \href {https://ui.adsabs.harvard.edu/abs/2021MNRAS.502.4328R} {502, 4328}

\bibitem[\protect\citeauthoryear{{Scodeggio} et~al.,}{{Scodeggio}
  et~al.}{2018}]{scodeggio18}
{Scodeggio} M.,  et~al., 2018, \mn@doi [\aap] {10.1051/0004-6361/201630114},
  \href {https://ui.adsabs.harvard.edu/abs/2018A&A...609A..84S} {609, A84}

\bibitem[\protect\citeauthoryear{{Serrano} et~al.,}{{Serrano}
  et~al.}{2023}]{Serrano:2023}
{Serrano} S.,  et~al., 2023, \mn@doi [\mnras] {10.1093/mnras/stad1399}, \href
  {https://ui.adsabs.harvard.edu/abs/2023MNRAS.523.3287S} {523, 3287}

\bibitem[\protect\citeauthoryear{{Smith}, {Cole}, {Baugh}, {Zheng}, {Angulo},
  {Norberg}  \& {Zehavi}}{{Smith} et~al.}{2017}]{smith17}
{Smith} A.,  {Cole} S.,  {Baugh} C.,  {Zheng} Z.,  {Angulo} R.,  {Norberg} P.,
   {Zehavi} I.,  2017, \mn@doi [\mnras] {10.1093/mnras/stx1432}, \href
  {https://ui.adsabs.harvard.edu/abs/2017MNRAS.470.4646S} {470, 4646}

\bibitem[\protect\citeauthoryear{{Soo} et~al.,}{{Soo} et~al.}{2021}]{Soo:2021}
{Soo} J. Y.~H.,  et~al., 2021, \mn@doi [\mnras] {10.1093/mnras/stab711}, \href
  {https://ui.adsabs.harvard.edu/abs/2021MNRAS.503.4118S} {503, 4118}

\bibitem[\protect\citeauthoryear{{Springel}, {White}, {Tormen}  \&
  {Kauffmann}}{{Springel} et~al.}{2001}]{Springel:2001}
{Springel} V.,  {White} S. D.~M.,  {Tormen} G.,   {Kauffmann} G.,  2001,
  \mn@doi [\mnras] {10.1046/j.1365-8711.2001.04912.x}, \href
  {https://ui.adsabs.harvard.edu/abs/2001MNRAS.328..726S} {328, 726}

\bibitem[\protect\citeauthoryear{{Springel} et~al.,}{{Springel}
  et~al.}{2005}]{Springel:2005}
{Springel} V.,  et~al., 2005, \mn@doi [\nat] {10.1038/nature03597}, \href
  {https://ui.adsabs.harvard.edu/abs/2005Natur.435..629S} {435, 629}

\bibitem[\protect\citeauthoryear{{Stothert} et~al.,}{{Stothert}
  et~al.}{2018}]{stothert18}
{Stothert} L.,  et~al., 2018, \mn@doi [\mnras] {10.1093/mnras/sty2491}, \href
  {https://ui.adsabs.harvard.edu/abs/2018MNRAS.481.4221S} {481, 4221}

\bibitem[\protect\citeauthoryear{{Strateva} et~al.,}{{Strateva}
  et~al.}{2001}]{Strateva:2001}
{Strateva} I.,  et~al., 2001, \mn@doi [\aj] {10.1086/323301}, \href
  {https://ui.adsabs.harvard.edu/abs/2001AJ....122.1861S} {122, 1861}

\bibitem[\protect\citeauthoryear{{Taylor} et~al.,}{{Taylor}
  et~al.}{2011}]{Taylor:2011}
{Taylor} E.~N.,  et~al., 2011, \mn@doi [\mnras]
  {10.1111/j.1365-2966.2011.19536.x}, \href
  {https://ui.adsabs.harvard.edu/abs/2011MNRAS.418.1587T} {418, 1587}

\bibitem[\protect\citeauthoryear{{Zehavi} et~al.,}{{Zehavi}
  et~al.}{2011}]{zehavi11}
{Zehavi} I.,  et~al., 2011, \mn@doi [\apj] {10.1088/0004-637X/736/1/59}, \href
  {https://ui.adsabs.harvard.edu/abs/2011ApJ...736...59Z} {736, 59}

\bibitem[\protect\citeauthoryear{{van den Busch} et~al.,}{{van den Busch}
  et~al.}{2020}]{errors:2020}
{van den Busch} J.~L.,  et~al., 2020, \mn@doi [\aap]
  {10.1051/0004-6361/202038835}, \href
  {https://ui.adsabs.harvard.edu/abs/2020A&A...642A.200V} {642, A200}

\makeatother
\end{thebibliography}

\begin{appendix}

\section{Interpolation scheme for apparent magnitudes}
\label{app:0}

\begin{figure}
   \centering
   \centering
    {\includegraphics[width=0.48\textwidth]{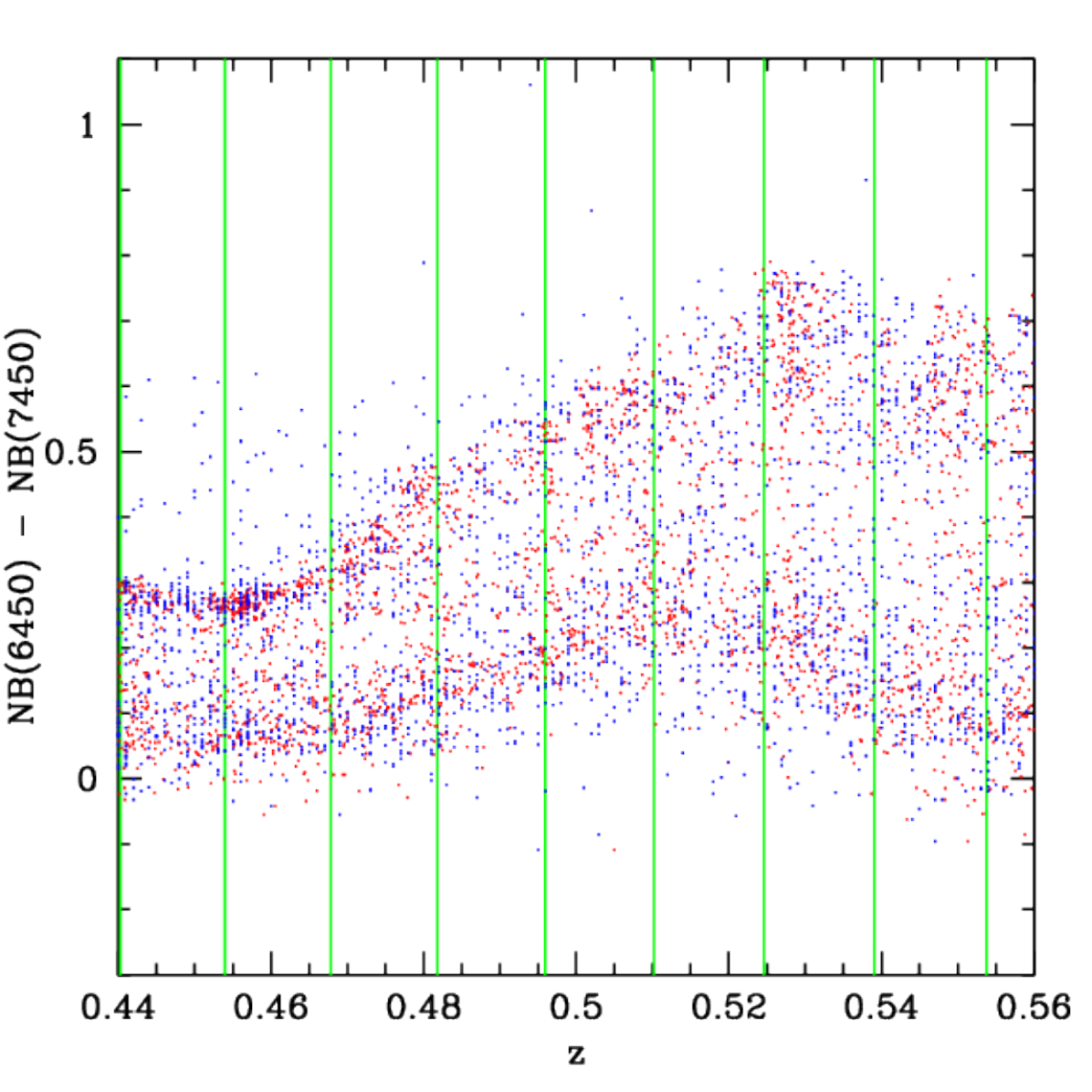} }%
     \caption{A colour - redshift plot as a test of the interpolation scheme for observer frame magnitudes using narrow band filters. The narrow band filters used have central wavelengths of 6450 and 7450 \AA\, as labelled in the $y$-axis. The \galf magnitudes are plotted without errors for this purpose. The red points show the galaxies plotted using the redshift from the lightcone. The blue points are plotted at the estimated photometric redshifts. The vertical green lines show the redshifts of the outputs in the P-Millennium N-body simulation. There is no stepping or banding apparent in the observed galaxy colour, even when plotted using the estimated photometric redshift. Furthermore, there is no indication that the estimated photometric redshifts favour the snapshot redshifts.}
     \label{fig:nb_zz}
\end{figure}

In the \galf model, the observer frame magnitude in a given band is defined at the redshift of each of the output snapshots of the simulation. 
The apparent magnitude of the galaxy at the redshift of lightcone crossing is calculated by interpolating in redshift between the observed magnitudes at the snapshots on either side of the lightcone crossing redshift \citep{Blaizot:2005,KW2007,merson13}.
\cite{merson13} showed that this scheme resulted in a smooth colour redshift distribution, matching the general form observed. 

In this section, we extend this test of the observer frame magnitude interpolation in two ways. 
First, we consider the colour - redshift relation for a colour defined using narrow band filters rather than broad band filters. 
The red points in Fig.~\ref{fig:nb_zz} show the \galf galaxy colours, without any photometric errors, plotted against their lightcone redshift. 
The vertical green lines mark the redshifts of the simulation snapshots. There is no stepping or discreteness visible in the red points. 
Next, we investigate the photometric redshifts estimated using the \galf galaxy photometry. 
These results are shown by the blue points in Fig.~\ref{fig:nb_zz}. Again, there is no preference for the photometric redshift code to return the N-body simulation snapshots, which suggests that any errors introduced by the interpolation scheme are smaller than those resulting from the redshift estimation. 

\section{Adding errors to the mock galaxy photometry}
\label{app:a}

\begin{table}

 \centering
 \caption{The narrow band magnitude limits for the W3 field, for a point source detected at $5 \sigma$. The first column gives the central wavelength of the filter in nm and the second column gives the magnitude limit for the band.}
	\label{tab:NBmaglim}
	\begin{tabular}{ll} 
		\hline
  $\lambda(\textrm{nm})$ & $m_{\textrm{lim}}$ \\
  \hline

   455 & 23.17 \\
   465 & 23.09 \\
   475 & 23.24 \\
   485 & 23.33 \\
   495 & 23.32 \\
   505 & 23.25 \\
   515 & 23.15 \\
   525 & 23.22 \\
   535 & 23.31  \\
   545 & 23.27 \\
   555 & 23.02 \\
   565 & 23.16 \\
   575 & 23.32 \\
   585 & 23.17 \\
   595 & 23.04 \\
   605 & 23.18 \\
   615 & 23.27 \\
   625 & 23.14 \\
   635 & 23.20 \\
   645 & 23.32 \\
   655 & 23.29 \\
   665 & 23.35 \\
   675 & 23.25 \\
   685 & 23.09 \\
   695 & 23.03 \\
   705 & 23.15 \\
   715 & 23.20 \\
   725 & 22.95 \\
   735 & 22.86 \\
   745 & 23.00 \\
   755 & 22.81 \\
   765 & 22.70 \\
   775 & 22.72 \\
   785 & 22.65 \\
   795 & 22.68 \\
   805 & 22.81 \\
   815 & 22.95 \\
   825 & 22.75 \\
   835 & 22.50 \\
   845 & 22.57 \\

		\hline
	\end{tabular}
\end{table}

We follow the method set out in \cite{errors:2020} to add errors to the magnitudes predicted for galaxies in the mock catalogue which reflect the observing strategy for PAUS. 
The errors are assumed to have a Gaussian distribution in magnitude. 
The perturbed magnitude in the band labelled by $j$, $m^{\rm obs}_{j}$, is obtained by adding a Gaussian distributed quantity, $x$, which has zero mean and variance $\sigma_{j}$ to the true magnitude predicted by {\tt GALFORM}, $m^{\rm true}_{j}$:  
\begin{equation}
    m^{\rm obs}_{j} = m^{\rm true}_{j} + x.
\end{equation}
The variance of the Gaussian is related to the signal-to-noise ratio in band $j$, $\left(S/N\right)_{j}$, by 
\begin{equation}
\sigma^{2}_{j} = \frac{2.5}{\ln 10} \frac{1}{(S/N)_{j}}.
\end{equation}

\cite{errors:2020} model the signal-to-noise ratio as a function of the magnitude limit in band $j$, $m^{\rm lim}_{j}$ as: 
\begin{equation}
\left(S/N\right)_{j} = 10^{-0.4(m^{\rm true}_{j} - m^{\rm lim}_{j})} \, f  \, k ,
\label{eq:sn}
\end{equation}
where $f$ is a factor which depends on the size of the galaxy if it is resolved and $k$ gives the signal-to-noise ratio for a point source at the magnitude limit. 
For an extended source, $f<1$. 
Here we assume $f=1$ for all sources and $k=5$, which means that all galaxies are treated as point sources and are detected with $S/N=5$ at the magnitude limit of the band in question. 
The magnitude limits in the broad band (BB) filters come from the CFHTLenS photometric catalogues for the W1 and W3 fields \citep{Erben:2013}. 
The PAUS narrow band magnitude limits correspond to $5 \sigma$ limits for a point source (see Table~\ref{tab:NBmaglim}).  
The estimation of the NB errors is described in \cite{Serrano:2023}, and takes into account the Poisson error in the electron count from the CCDs and the sky noise in the aperture. Note that \galf makes a prediction of the size of the disk and bulge component of each galaxy, so in principle, we could have applied a more accurate model for the photometric errors, which took into account whether or not the galaxy is an extended source. 
However, the predictions for the sizes of disks and bulges are some of the less accurate \galf predictions (see, for example, the galaxy size - luminosity plots in \citealt{lacey:2016} and \citealt{elliott21}). 
Hence in the simple model for photometric errors presented here, we have forced the assumption that \textit{all} model galaxies are point sources. 
The results recovered in Fig.~\ref{fig:zb_zspec} reassure us that this methodology for simulating errors on the magnitudes is accurate enough to get photometric redshifts and photometric redshift errors in agreement with the observations, as described in the following Appendix~\ref{app:b}.

\section{The characteristics of photometric redshifts in the mock catalogue}
\label{app:b}

After assigning magnitude errors to the model galaxies as described above in Appendix~\ref{app:a}, we run the BCNz2 code developed by \cite{eriksen19} to estimate photometric redshifts for the mock catalogue. Here we examine the resulting scatter in the estimated redshift and the fraction of outliers, i.e. redshifts with catastrophic errors, and compare these to the results found for the observations. 

Following \cite{Brammer:2008} a quality factor, $Qz$ is calculated for each photometric redshift to quantify our confidence in the accuracy of the photometric redshift: 
\begin{equation}
Qz = \frac{\chi^{2}}{{N_{\rm f}-3}}\left(\frac{ z^{99} - z^{1} }{\rm{ODDS}(\Delta z = 0.0035)} \right),
\end{equation}
where $N_{\rm f}$ is the number of filters used to sample the spectral energy distribution (SED) of the galaxy, $\chi^2$ is the metric describing how well the template SED fits the observations, $z^{99}$ is the redshift below which 99 per cent of the redshift probability distribution lies and $z^{1}$ is the redshift below which 1 per cent of the probability density function lies. The $\rm{ODDS}$ quantity is defined as 
\begin{equation}
{\rm ODDS} = \int_{z_{\rm b} - \Delta z}^{z_{\rm b}+ \Delta z}
p(z) {\rm d} z, 
\end{equation}
where $p(z)$ is the redshift probability density function and $z_{\rm b}$ is the mode of $p(z)$. 
Note that $\Delta z= 0.0035$ is smaller than the value typically used for BB filters, and has been reduced to reflect the width of the PAUS NB filters. 
These choices are discussed at length in \cite{eriksen19}. 
A galaxy with a good photometric redshift quality has a \textit{low} $Qz$ value as this implies a low value of $\chi^{2}$ and a high value for the $\rm{ODDS}$ (due to a peaked, narrow $p(z)$). 
Calculating a $Qz$ value for the model galaxies allows us to study the errors and metrics for different subsamples of galaxies, as is usually done for the data.  

\begin{figure}
   \centering
   \centering
    {\includegraphics[width=0.48\textwidth]{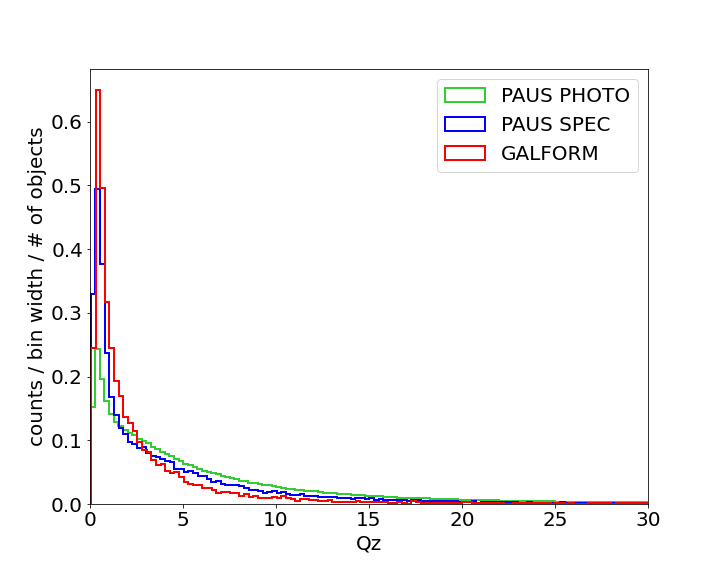} }%
    {\includegraphics[width=0.48\textwidth]{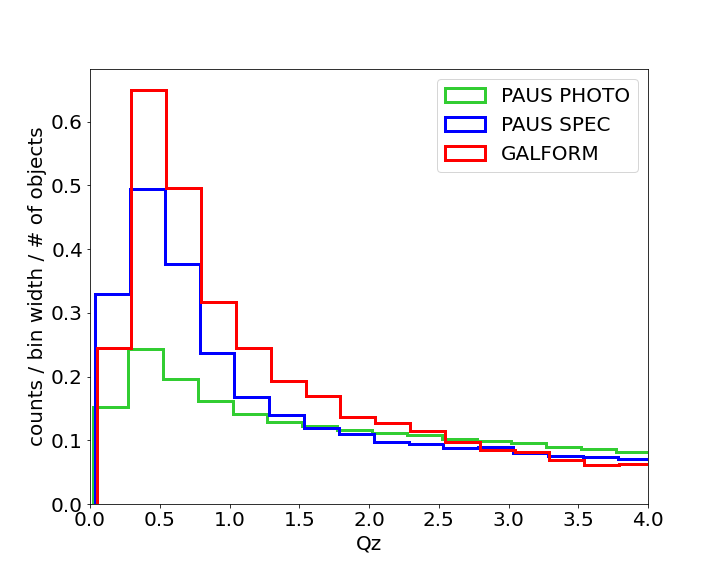} }
     \caption{Normalised distribution of the redshift quality factor $Qz$ for three different samples: the red histogram shows the galform lightcone model, the blue histogram shows a spectroscopically matched PAUS subsample (PAUS SPEC) and the green histogram for the full photometric PAUS sample used here (PAUS PHOTO). The lower panel is a zoomed-in version of the upper panel showing the low $Qz$ or good photometric redshift region.}
     \label{fig:qz_histo}
\end{figure}

The distribution of $Qz$ values recovered from the mock catalogue is compared with that from the observations in Fig.~\ref{fig:qz_histo} (the different panels have different levels of zooming). 
The distribution for the mock galaxies is impressively close to those estimated for the PAUS galaxies, particularly for the subsample with spectroscopic matches (labelled as PAUS SPEC\footnote{The PAUS SPEC is a subsample of PAUS PHOTO that has been matched with some overlapping spectroscopic surveys, in order to have spectroscopic redshift. This subsample has been already used in Section~\ref{sec:photoz} to obtain estimates of the photometric redshift errors (see Fig.~\ref{fig:zb_zspec}).}).  

\begin{figure}
   \centering
   \centering
    {\includegraphics[width=0.48\textwidth]{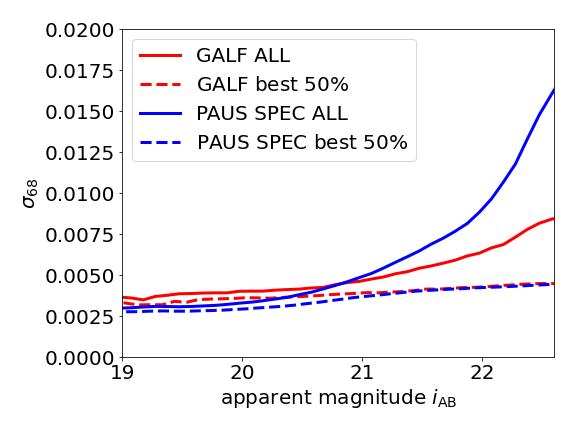} }%
    {\includegraphics[width=0.48\textwidth]{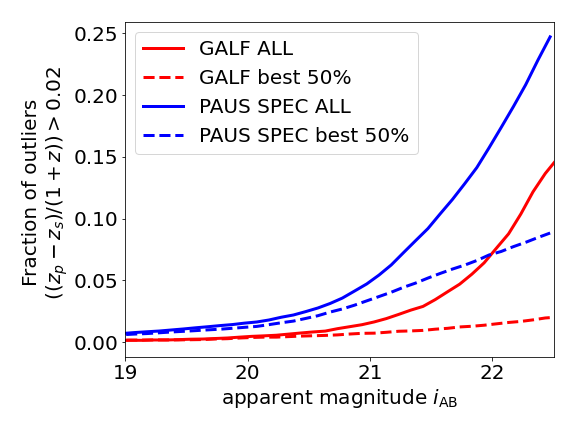} }
     \caption{Upper panel: cumulative plot of $\sigma_{68}$ as a function of magnitude. Red lines show the scatter obtained from the {\tt GALFORM} mock while the blue lines are from the spectroscopically matched PAUS subsample (PAUS SPEC). Dashed lines are for the best $50$ per cent of the relative sample based on the $Qz$ value.
     Lower panel: cumulative fraction of outliers as a function of magnitude. The red continuous lines show the results for the {\tt GALFORM} mock (whereas the dashed line shows the outlier fraction for the 50 per cent highest quality redshifts based on the $Qz$ value). Blue lines show the corresponding quantities for the spectroscopically matched PAUS sample (continuous lines are for the full sample and the dashed lines for the best 50 per cent). Both panels can be compared with fig.~3 of \protect\cite{eriksen19}.}
     \label{fig:sn}
\end{figure}

We next consider, in Fig.~\ref{fig:sn}, the centralised estimate of the scatter, $\sigma_{68}$, as a function of magnitude, for different subsamples from the mock, defined using the $Qz$ value. 
This plot can be compared with the upper panel of fig.~3 from \cite{Eriksen2020}. 
The agreement with the PAUS SPEC sample is remarkably good (especially when selecting the best 50 per cent of the sample based on the $Qz$ value), though we note that this sample is biased towards brighter galaxies than the magnitude-limited mock catalogue, because of the difficulties in getting spectra of very faint objects.  

As well as the scatter, the performance of the photometric redshift estimation can be quantified using the fraction of outliers produced. 
Following \cite{eriksen19}, we define the fraction of outliers as the number of galaxies, normalised by the total number of galaxies in the sample, that satisfy:
\begin{equation}
    \label{eq:outliers}
    \frac{|z_{\rm photo} - z_{\rm spec}|}{(1 + z_{\rm spec})} > 0.02. 
\end{equation}
The outlier fraction is shown in the lower panel of Fig.~\ref{fig:sn}. The mock shows a similar trend to the PAUS data for the outlier fraction as a function of magnitude, but with the overall values being somewhat lower in the mock than in the observations. This holds true both for the full sample (solid lines) and the 50 per cent best quality redshift according to the $Qz$ criterion (dashed lines).  
\label{lastpage}

\end{appendix}
\end{document}